\documentclass[11pt]{article}

\usepackage{amsmath}

\usepackage{color,epsfig,graphics,amsfonts,amsmath}
\numberwithin{equation}{section}

\def\ba{\begin{array}}
\def\ea{\end{array}}

\begin{document}

\title{E. Ovsiyuk, O. Veko, M. Neagu, \\ V. Balan, V. Red'kov \\[3mm]
Elementary constituents of the group $SL(4,\mathbb{R})$, \\
and
classification of the Mueller matrices }

\date{}

\maketitle

\begin{abstract}

 The  goal of this
paper is to develop a systematic method of locating the Mueller
matrices within the class of the matrices of the real group
$SL(4,\mathbb{R})$. The main idea is
to construct the general transformation of the group $SL(4,%
\mathbb{R})$ (whose real matrices have unit determinant) is
straightforward, but to analyze the adequacy of such a
transformation for describing Mueller matrices is highly
nontrivial. However, using the technique of Dirac matrices, we can
quite easy and explicitly describe all the 16 one-parametric
subgroups, from which, using the all possible products, emerges
the whole group $SL(4,\mathbb{R})$. As a matter of fact, for these
separate 1-parametric subgroups the question of their adequacy of
describing Mueller matrices becomes sufficiently simple and thus
we  obtain in each case a definite answer.

\end{abstract}

%=========================================================

%\thispagestyle{empty}

\section{Elementary 1-parametric generators of the group $GL(4,\mathbb{R})$}

It is well known the cornerstone role played in polarization optics by the
Mueller matrices \cite{1}. Distinguished subsets of the set of Mueller
matrices generate group structures, which are isomorphic to the group of
rotations or to the Lorentz group.

Since the Mueller matrices are real, of order $4\times 4$, and acting on the
4-dimensional real Stokes vector, to investigate the sets of all possible
Mueller matrices, one can use the parametrization of 4-dimensional matrices
which is obtained on the ground of using the basis of Dirac matrices,
developed in the works \cite{2,3,4}. The main goal of this paper is to
develop a systematic method of locating the Mueller matrices within the
class of the matrices of the real group $SL(4,\mathbb{R})$. The main idea is
the following. To construct the general transformation of the group $SL(4,%
\mathbb{R})$ (whose real matrices have unit determinant) is
straightforward, but to analyze the adequacy of such a
transformation for describing Mueller matrices is highly
nontrivial (practically impossible). However, using the technique
of Dirac matrices, we can quite easy and explicitly describe all
the 16 one-parametric subgroups, from which, using the all
possible products, emerges the whole group $SL(4,\mathbb{R})$. As
a matter of fact, for these separate 1-parametric subgroups the
question of their adequacy of describing Mueller matrices becomes
sufficiently simple and thus we have reasons to expect in each
case a definite answer.

The explicit form of the Dirac 16-dimensional basis (using the Weyl spinor
representation) \cite{2,3,4} is:
\begin{eqnarray}
\;\qquad \gamma ^{5} &=&\left\vert
\begin{array}{cccc}
-1 & 0 & 0 & 0 \\
0 & -1 & 0 & 0 \\
0 & 0 & 1 & 0 \\
0 & 0 & 0 & 1%
\end{array}%
\right\vert ,\text{ }\gamma ^{0}=\left\vert
\begin{array}{cccc}
0 & 0 & 1 & 0 \\
0 & 0 & 0 & 1 \\
1 & 0 & 0 & 0 \\
0 & 1 & 0 & 0%
\end{array}%
\right\vert ,\text{ }i\gamma ^{5}\gamma ^{0}=\left\vert
\begin{array}{cccc}
0 & 0 & -i & 0 \\
0 & 0 & 0 & -i \\
i & 0 & 0 & 0 \\
0 & i & 0 & 0%
\end{array}%
\right\vert ,  \notag \\
\qquad i\gamma ^{1} &=&\left\vert
\begin{array}{cccc}
0 & 0 & 0 & -i \\
0 & 0 & -i & 0 \\
0 & i & 0 & 0 \\
i & 0 & 0 & 0%
\end{array}%
\right\vert ,\text{ }\gamma ^{5}\gamma ^{1}=\left\vert
\begin{array}{cccc}
0 & 0 & 0 & 1 \\
0 & 0 & 1 & 0 \\
0 & 1 & 0 & 0 \\
1 & 0 & 0 & 0%
\end{array}%
\right\vert ,\text{ }i\gamma ^{2}=\left\vert
\begin{array}{cccc}
0 & 0 & 0 & -1 \\
0 & 0 & 1 & 0 \\
0 & 1 & 0 & 0 \\
-1 & 0 & 0 & 0%
\end{array}%
\right\vert ,  \notag \\
\gamma ^{5}\gamma ^{2} &=&\left\vert
\begin{array}{cccc}
0 & 0 & 0 & -i \\
0 & 0 & i & 0 \\
0 & -i & 0 & 0 \\
i & 0 & 0 & 0%
\end{array}%
\right\vert ,\text{ }i\gamma ^{3}=\left\vert
\begin{array}{cccc}
0 & 0 & -i & 0 \\
0 & 0 & 0 & i \\
i & 0 & 0 & 0 \\
0 & -i & 0 & 0%
\end{array}%
\right\vert ,\text{ }\gamma ^{5}\gamma ^{3}=\left\vert
\begin{array}{cccc}
0 & 0 & 1 & 0 \\
0 & 0 & 0 & -1 \\
1 & 0 & 0 & 0 \\
0 & -1 & 0 & 0%
\end{array}%
\right\vert ,  \notag \\
2\sigma ^{01} &=&\left\vert
\begin{array}{cccc}
0 & 1 & 0 & 0 \\
1 & 0 & 0 & 0 \\
0 & 0 & 0 & -1 \\
0 & 0 & -1 & 0%
\end{array}%
\right\vert ,\text{ }2\sigma ^{02}=\left\vert
\begin{array}{cccc}
0 & -i & 0 & 0 \\
i & 0 & 0 & 0 \\
0 & 0 & 0 & i \\
0 & 0 & -i & 0%
\end{array}%
\right\vert ,\text{ }2\sigma ^{03}=\left\vert
\begin{array}{cccc}
1 & 0 & 0 & 0 \\
0 & -1 & 0 & 0 \\
0 & 0 & -1 & 0 \\
0 & 0 & 0 & 1%
\end{array}%
\right\vert ,  \notag \\
2i\sigma ^{12} &=&\left\vert
\begin{array}{cccc}
1 & 0 & 0 & 0 \\
0 & -1 & 0 & 0 \\
0 & 0 & 1 & 0 \\
0 & 0 & 0 & -1%
\end{array}%
\right\vert ,\text{ }2i\sigma ^{23}=\left\vert
\begin{array}{cccc}
0 & 1 & 0 & 0 \\
1 & 0 & 0 & 0 \\
0 & 0 & 0 & 1 \\
0 & 0 & 1 & 0%
\end{array}%
\right\vert ,\text{ }2i\sigma ^{31}=\left\vert
\begin{array}{cccc}
0 & -i & 0 & 0 \\
i & 0 & 0 & 0 \\
0 & 0 & 0 & -i \\
0 & 0 & i & 0%
\end{array}%
\right\vert .  \notag
\end{eqnarray}

All these 15 matrices (let us note them as $\Lambda _{k}$) are of Gell-Mann
type: - this means they have null trace and they are Hermitian too;
moreover, the square of each of them is equal to $I$:
\begin{equation*}
\mbox{Sp}\Lambda =0,\qquad (\Lambda )^{2}=I,\qquad (\Lambda )^{+}=\Lambda
,\qquad \Lambda \in \{\Lambda _{k}:k=1,...,15\;\}.
\end{equation*}

Due to the identity $(\Lambda )^{2}=I$, the exponential of each from these
matrices $\Lambda $ is
\begin{equation*}
U=e^{ia\Lambda }=\cos a+i\sin a\Lambda ,\qquad \mbox{det}\;e^{ia\Lambda
}=+1,\qquad a\in \mathbb{R}.
\end{equation*}

We will use the following notations \cite{2, 3} for some specially chosen
six generators $\Lambda _{i}$:
\begin{eqnarray}
&&%
\begin{array}{lll}
\alpha _{1}=\gamma ^{0}\gamma ^{2}, & \alpha _{2}=i\gamma ^{0}\gamma ^{5}, &
\alpha _{3}=\gamma ^{5}\gamma ^{2},\medskip \\
\alpha _{i}^{2}=I, & \alpha _{1}\alpha _{2}=i\alpha _{3}, & \alpha
_{2}\alpha _{1}=-i\alpha _{3};%
\end{array}
\label{1.3} \\
&&%
\begin{array}{lll}
\beta _{1}=i\gamma ^{3}\gamma ^{1}, & \beta _{2}=i\gamma ^{3}, & \beta
_{3}=i\gamma ^{1},\medskip \\
\beta _{i}^{2}=I, & \beta _{1}\beta _{2}=i\beta _{3}, & \beta _{2}\beta
_{1}=-i\beta _{3}.%
\end{array}
\label{1.4}
\end{eqnarray}

Note that we have the commutation relations $\alpha _{j}\beta _{k}=\beta
_{k}\alpha _{j}$. Due to this commutation one can construct nine Abelian
2-parametric subgroups. For instance, $e^{ia_{1}\alpha _{1}}e^{ib_{1}\beta
_{1}}=e^{ib_{1}\beta _{1}}e^{ia_{1}\alpha _{1}}\;$and so on$.$

All the possible products of matrices (\ref{1.4}) provide us nine Lie
algebra generators:
\begin{eqnarray}
\left.
\begin{array}{lll}
A_{1}= \alpha_{1} \beta_{1}= -\gamma^{5} , & B_{1}= \alpha_{1} \beta_{2}=
\gamma^{5}\gamma^{1}, & C_{1}=\alpha_{1} \beta_{3}=\gamma^{3} \gamma^{5}, \\%
[2mm]
A_{2}= \alpha_{2} \beta_{1}=-i \gamma^2, & B_{2}= \alpha_{2} \beta_{2}=-i
\gamma^{1}\gamma^2, & C_{2}= \alpha_{2} \beta_{3}=-i \gamma^2\gamma^{3}, \\%
[2mm]
A_{3}= \alpha_{3} \beta_{1}= \gamma^{0}, & B_{3}= \alpha_{3} \beta_{2}=
\gamma^{0}\gamma^{1}, & C_{3}= \alpha_{3} \beta_{3}= \gamma^{0}\gamma^{3} .%
\end{array}
\right.  \label{1.5}
\end{eqnarray}

We further specify the explicit form of the 15 elementary 1-parametric
unitary transformations:
\begin{eqnarray}
\alpha _{1} &=&\left\vert
\begin{array}{cc}
\sigma _{2} & 0 \\
0 & -\sigma _{2}%
\end{array}%
\right\vert ,\qquad \alpha _{2}=\left\vert
\begin{array}{cc}
0 & iI_2 \\
-iI_2 & 0%
\end{array}%
\right\vert ,\qquad \alpha _{3}=\left\vert
\begin{array}{cc}
0 & \sigma _{2} \\
\sigma _{2} & 0%
\end{array}%
\right\vert ,  \notag \\
U_{1}^{\alpha } &=&\left\vert
\begin{array}{cc}
\cos \phi +i\sin \phi \sigma _{2} & 0 \\
0 & \cos \phi -i\sin \phi \sigma _{2}%
\end{array}%
\right\vert ,  \notag \\
U_{2}^{\alpha } &=&\left\vert
\begin{array}{cc}
\cos \phi & -\sin \phi \\
\sin \phi & \cos \phi%
\end{array}%
\right\vert ,\qquad U_{3}^{\alpha }=\left\vert
\begin{array}{cc}
\cos \phi & i\sin \phi \sigma _{2} \\
i\sin \phi \sigma _{2} & \cos \phi%
\end{array}%
\right\vert ;  \label{1.6a}
\end{eqnarray}%
\begin{eqnarray}
\beta _{1} &=&\left\vert
\begin{array}{cc}
\sigma _{2} & 0 \\
0 & \sigma _{2}%
\end{array}%
\right\vert ,\qquad \beta _{2}=\left\vert
\begin{array}{cc}
0 & -i\sigma _{3} \\
i\sigma _{3} & 0%
\end{array}%
\right\vert ,\qquad \beta _{3}=\left\vert
\begin{array}{cc}
0 & -i\sigma _{1} \\
i\sigma _{1} & 0%
\end{array}%
\right\vert ,  \notag \\
U_{1}^{\beta } &=&\left\vert
\begin{array}{cc}
\cos \phi +i\sin \phi \sigma _{2} & 0 \\
0 & \cos \phi +i\sin \phi \sigma _{2}%
\end{array}%
\right\vert ,  \notag \\
U_{2}^{\beta } &=&\left\vert
\begin{array}{cc}
\cos \phi & \sin \phi \sigma _{3} \\
-\sin \phi \sigma _{3} & \cos \phi%
\end{array}%
\right\vert ,\qquad U_{3}^{\beta }=\left\vert
\begin{array}{cc}
\cos \phi & \sin \phi \sigma _{1} \\
-\sin \phi \sigma _{1} & \cos \phi%
\end{array}%
\right\vert ;  \label{1.6b}
\end{eqnarray}%
\begin{eqnarray}
A_{1} &=&\left\vert
\begin{array}{cc}
I & 0 \\
0 & -I%
\end{array}%
\right\vert ,\qquad A_{2}=\left\vert
\begin{array}{cc}
0 & i\sigma _{2} \\
-i\sigma _{2} & 0%
\end{array}%
\right\vert ,\qquad A_{3}=\left\vert
\begin{array}{cc}
0 & I \\
I & 0%
\end{array}%
\right\vert ,  \notag \\
U_{1}^{A} &=&\left\vert
\begin{array}{cc}
\cos \phi +i\sin \phi & 0 \\
0 & \cos \phi -i\sin \phi%
\end{array}%
\right\vert ,\qquad  \notag \\
U_{2}^{A} &=&\left\vert
\begin{array}{cc}
\cos \phi & -\sin \phi \sigma _{2} \\
\sin \phi \sigma _{2} & \cos \phi%
\end{array}%
\right\vert ,\qquad U_{3}^{A}=\left\vert
\begin{array}{cc}
\cos \phi & i\sin \phi \\
i\sin \phi & \cos \phi%
\end{array}%
\right\vert ,  \label{1.6c}
\end{eqnarray}%
\begin{eqnarray}
B_{1} &=&\left\vert
\begin{array}{cc}
0 & \sigma _{1} \\
\sigma _{1} & 0%
\end{array}%
\right\vert ,\qquad B_{2}=\left\vert
\begin{array}{cc}
-\sigma _{3} & 0 \\
0 & -\sigma _{3}%
\end{array}%
\right\vert ,\qquad B_{3}=\left\vert
\begin{array}{cc}
-\sigma _{1} & 0 \\
0 & \sigma _{1}%
\end{array}%
\right\vert ,  \notag \\
U_{1}^{B} &=&\left\vert
\begin{array}{cc}
\cos \phi & i\sin \phi \sigma _{1} \\
i\sin \phi \sigma _{1} & \cos \phi%
\end{array}%
\right\vert ,  \notag \\
U_{2}^{B} &=&\left\vert
\begin{array}{cc}
\cos \phi -i\sin \phi \sigma _{3} & 0 \\
0 & \cos \phi -i\sin \phi \sigma _{3}%
\end{array}%
\right\vert ,  \notag \\
U_{3}^{B} &=&\left\vert
\begin{array}{cc}
\cos \phi -i\sin \phi \sigma _{1} & 0 \\
0 & \cos \phi +i\sin \phi \sigma _{1}%
\end{array}%
\right\vert ;  \label{1.6d}
\end{eqnarray}%
\begin{eqnarray}
C_{1} &=&\left\vert
\begin{array}{cc}
0 & -\sigma _{3} \\
-\sigma _{3} & 0%
\end{array}%
\right\vert ,\qquad C_{2}=\left\vert
\begin{array}{cc}
-\sigma _{1} & 0 \\
0 & -\sigma _{1}%
\end{array}%
\right\vert ,\qquad C_{3}=\left\vert
\begin{array}{cc}
\sigma _{3} & 0 \\
0 & -\sigma _{3}%
\end{array}%
\right\vert ,  \notag \\
U_{1}^{C} &=&\left\vert
\begin{array}{cc}
\cos \phi & -i\sin \phi \sigma _{3} \\
-i\sin \phi \sigma _{3} & \cos \phi%
\end{array}%
\right\vert ,  \notag \\
U_{2}^{C} &=&\left\vert
\begin{array}{cc}
\cos \phi -i\sin \phi \sigma _{1} & 0 \\
0 & \cos \phi -i\sin \phi \sigma _{1}%
\end{array}%
\right\vert ,  \notag \\
U_{3}^{C} &=&\left\vert
\begin{array}{cc}
\cos \phi +i\sin \phi \sigma _{3} & 0 \\
0 & \cos \phi -i\sin \phi \sigma _{3}%
\end{array}%
\right\vert ,  \label{1.6e}
\end{eqnarray}%
and hence we easily get the 15 real 1-parametric $4\times 4$-transformations
from the group $SL(4,\mathbb{R})$:
\begin{eqnarray}
U_{1}^{\alpha }(\phi ) &=&\left\vert
\begin{array}{cccc}
\cos \phi & \sin \phi & 0 & 0 \\
-\sin \phi & \cos \phi & 0 & 0 \\
0 & 0 & \cos \phi & -\sin \phi \\
0 & 0 & \sin \phi & \cos \phi%
\end{array}%
\right\vert ,  \notag \\
U_{2}^{\alpha }(\phi ) &=&\left\vert
\begin{array}{cccc}
\cos \phi & 0 & -\sin \phi & 0 \\
0 & \cos \phi & 0 & -\sin \phi \\
\sin \phi & 0 & \cos \phi & 0 \\
0 & \sin \phi & 0 & \cos \phi%
\end{array}%
\right\vert ,  \notag \\
U_{3}^{\alpha }(\phi ) &=&\left\vert
\begin{array}{cccc}
\cos \phi & 0 & 0 & \sin \phi \\
0 & \cos \phi & -\sin \phi & 0 \\
0 & \sin \phi & \cos \phi & 0 \\
-\sin \phi & 0 & 0 & \cos \phi%
\end{array}%
\right\vert ;  \label{1.7a}
\end{eqnarray}%
\begin{eqnarray}
U_{1}^{\beta }(\phi ) &=&\left\vert
\begin{array}{cccc}
\cos \phi & \sin \phi & 0 & 0 \\
-\sin \phi & \cos \phi & 0 & 0 \\
0 & 0 & \cos \phi & \sin \phi \\
0 & 0 & -\sin \phi & \cos \phi%
\end{array}%
\right\vert ,  \notag \\
U_{2}^{\beta }(\phi ) &=&\left\vert
\begin{array}{cccc}
\cos \phi & 0 & \sin \phi & 0 \\
0 & \cos \phi & 0 & -\sin \phi \\
-\sin \phi & 0 & \cos \phi & 0 \\
0 & \sin \phi & 0 & \cos \phi%
\end{array}%
\right\vert ,  \notag \\
U_{3}^{\beta }(\phi ) &=&\left\vert
\begin{array}{cccc}
\cos \phi & 0 & 0 & \sin \phi \\
0 & \cos \phi & \sin \phi & 0 \\
0 & -\sin \phi & \cos \phi & 0 \\
-\sin \phi & 0 & 0 & \cos \phi%
\end{array}%
\right\vert ;  \label{1.7b}
\end{eqnarray}%
\begin{eqnarray}
U_{1}^{A}(i\lambda ) &=&\left\vert
\begin{array}{cccc}
e^{-\lambda } & 0 & 0 & 0 \\
0 & e^{-\lambda } & 0 & 0 \\
0 & 0 & e^{\lambda } & 0 \\
0 & 0 & 0 & e^{\lambda }%
\end{array}%
\right\vert ,  \notag \\
U_{2}^{A}(i\beta ) &=&\left\vert
\begin{array}{cccc}
\cosh \;\beta & 0 & 0 & -\sinh \;\beta \\
0 & \cosh \;\beta & \sinh \;\beta & 0 \\
0 & \sinh \;\beta & \cosh \;\beta & 0 \\
-\sinh \;\beta & 0 & 0 & \cosh \;\beta%
\end{array}%
\right\vert ,  \notag \\
U_{3}^{A}(i\beta ) &=&\left\vert
\begin{array}{cccc}
\cosh \;\beta & 0 & -\sinh \;\beta & 0 \\
0 & \cosh \;\beta & 0 & -\sinh \;\beta \\
-\sinh \;\beta & 0 & \cosh \;\beta & 0 \\
0 & -\sinh \;\beta & 0 & \cosh \;\beta%
\end{array}%
\right\vert ,  \label{1.7c}
\end{eqnarray}%
\begin{eqnarray}
U_{1}^{B}(i\beta ) &=&\left\vert
\begin{array}{cccc}
\cosh \;\beta & 0 & 0 & -\sinh \;\beta \\
0 & \cosh \;\beta & -\sinh \;\beta & 0 \\
0 & -\sinh \;\beta & \cosh \;\beta & 0 \\
-\sinh \;\beta & 0 & 0 & \cosh \;\beta%
\end{array}%
\right\vert ,  \notag \\
U_{2}^{B}(i\lambda ) &=&\left\vert
\begin{array}{cccc}
e^{\lambda } & 0 & 0 &  \\
0 & e^{-\lambda } & 0 & 0 \\
0 & 0 & e^{\lambda } & 0 \\
0 & 0 & 0 & e^{-\lambda }%
\end{array}%
\right\vert ,  \notag \\
U_{3}^{B}(i\beta ) &=&\left\vert
\begin{array}{cccc}
\cosh \;\beta & \sinh \;\beta & 0 & 0 \\
\sinh \;\beta & \cosh \;\beta & 0 & 0 \\
0 & 0 & \cosh \;\beta & -\sinh \;\beta \\
0 & 0 & -\sinh \;\beta & \cosh \;\beta%
\end{array}%
\right\vert ;  \label{1.7d}
\end{eqnarray}%
\begin{eqnarray}
U_{1}^{C}(i\beta ) &=&\left\vert
\begin{array}{cccc}
\cosh \;\beta & 0 & \sinh \;\beta & 0 \\
0 & \cosh \;\beta & 0 & -\sinh \;\beta \\
\sinh \;\beta & 0 & \cosh \;\beta & 0 \\
0 & -\sinh \;\beta & 0 & \cosh \;\beta%
\end{array}%
\right\vert ,  \notag \\
U_{2}^{C}(i\beta ) &=&\left\vert
\begin{array}{cccc}
\cosh \;\beta & \sinh \;\beta & 0 & 0 \\
\sinh \;\beta & \cosh \;\beta & 0 & 0 \\
0 & 0 & \cosh \;\beta & \sinh \;\beta \\
0 & 0 & \sinh \;\beta & \cosh \;\beta%
\end{array}%
\right\vert ,  \notag \\
U_{3}^{C}(i\lambda ) &=&\left\vert
\begin{array}{cccc}
e^{-\lambda } & 0 & 0 &  \\
0 & e^{\lambda } & 0 & 0 \\
0 & 0 & e^{\lambda } & 0 \\
0 & 0 & 0 & e^{-\lambda }%
\end{array}%
\right\vert .  \label{1.7e}
\end{eqnarray}

Note that to the generator $\lambda _{0}=I$ corresponds the finite element%
\begin{equation*}
U_{0}(i\lambda )=e^{-\lambda }\left\vert
\begin{array}{cccc}
1 & 0 & 0 & 0 \\
0 & 1 & 0 & 0 \\
0 & 0 & 1 & 0 \\
0 & 0 & 0 & 1%
\end{array}%
\right\vert .
\end{equation*}

\section{On the (pseudo) Euclidean rotations of Stokes 4-vectors}

A real $4\times 4$-matrix $M$ may be considered as being of Mueller type and
acting on polarized light (completely or partially) if $M_{ab}S_{a}=S_{a}^{%
\prime }$:%
\begin{eqnarray}
S_{0}^{\prime } &=&M_{00}S_{0}+M_{01}S_{1}+M_{02}S_{2}+M_{03}S_{3},  \notag
\\
S_{1}^{\prime } &=&M_{10}S_{0}+M_{11}S_{1}+M_{12}S_{2}+M_{13}S_{3},  \notag
\\
S_{2}^{\prime } &=&M_{20}S_{0}+M_{21}S_{1}+M_{22}S_{2}+M_{23}S_{3},  \notag
\\
S_{3}^{\prime } &=&M_{30}S_{0}+M_{31}S_{1}+M_{32}S_{2}+M_{33}S_{3},  \notag
\end{eqnarray}%
where the following two inequalities hold:
\begin{eqnarray}
S_{0} &\geq &0,\qquad S^{2}\geq 0,\qquad (S^{2}\equiv
S_{0}^{2}-S_{1}^{2}-S_{2}^{2}-S_{3}^{2}),  \label{2.1a} \\
S_{0}^{\prime } &\geq &0,\qquad S^{\prime 2}\geq 0,\qquad (S^{\prime
2}\equiv S_{0}^{\prime 2}-S_{1}^{\prime 2}-S_{2}^{\prime 2}-S_{3}^{\prime
2}).  \label{2.1b}
\end{eqnarray}

In more detailed form, these inequalities look as%
\begin{equation}
\begin{array}{l}
M_{00}S_{0}+M_{01}S_{1}+M_{02}S_{2}+M_{03}S_{3}\geq 0,\medskip \\
(M_{00}S_{0}+M_{01}S_{1}+M_{02}S_{2}+M_{03}S_{3})^{2}-(M_{10}S_{0}+M_{11}S_{1}+M_{12}S_{2}+M_{13}S_{3})^{2}-\medskip
\\
(M_{20}S_{0}+M_{21}S_{1}+M_{22}S_{2}+M_{23}S_{3})^{2}-(M_{30}S_{0}+M_{31}S_{1}+M_{32}S_{2}+M_{33}S_{3})^{2}\geq 0.%
\end{array}
\label{2.1c}
\end{equation}%
We recall that $S_{0}$ is the intensity $I$ of the light beam, and $%
S_{i}=S_{0}p_{i}=Ip_{i}$, where $p_{i}$ is the polarization vector.
Correspondingly, the above inequalities read as%
\begin{equation}
\begin{array}{l}
p_{1}^{2}+p_{2}^{2}+p_{3}^{2}\leq 1,\medskip \qquad
M_{00}+M_{01}p_{1}+M_{02}p_{2}+M_{03}p_{3}\geq 0, \\
(M_{00}+M_{01}p_{1}+M_{02}p_{2}+M_{03}p_{3})^{2}-(M_{10}+M_{11}p_{1}+M_{12}p_{2}+M_{13}p_{3})^{2}-\medskip
\\
(M_{20}+M_{21}p_{1}+M_{22}p_{2}+M_{23}p_{3})^{2}-(M_{30}+M_{31}p_{1}+M_{32}p_{2}+M_{33}p_{3})^{2}\geq 0.%
\end{array}
\label{2.1d}
\end{equation}

The problem we face here is very complicated due to the big number of the
independent parameters -- elements of the matrix $M_{ab}$. So, we may expect
various solutions. In the first place, we are interested in Mueller matrix
sets which exhibit a group structure.

The most evident and known such sets are the 3-parametric group of the
Euclidean 3-rotations, and the 6-parametric group of the pseudo-Euclidean
rotations, which form the Lorentz group. We shall describe them below.
Moreover, we shall consider all the 1-parametric elementary generators of
the real linear group $SL(4,\mathbb{R})$. Examination of other subgroups is
a subject for further concern.

Let us consider, by using the notation introduced above, the
Lorentzian rotations of the Stokes 4-vectors. To this end let us
specify  the following
two subgroups:%
\begin{eqnarray}
R_{\alpha }(k_{0},k_{i}) &=&k_{0}\;I+k_{i}\alpha ^{i}=\left\vert
\begin{array}{rrrr}
k_{0} & k_{1} & k_{2} & k_{3} \\
-k_{1} & k_{0} & -k_{3} & k_{2} \\
-k_{2} & k_{3} & k_{0} & -k_{1} \\
-k_{3} & -k_{2} & k_{1} & k_{0}%
\end{array}%
\right\vert ,  \notag \\
k_{0}^{\prime \prime } &=&k_{0}^{\prime }\;k_{0}-k_{i}^{\prime
}\;k_{i},\qquad k_{n}^{\prime \prime }=k_{0}^{\prime }\;k_{n}+k_{n}^{\prime
}\;k_{0}+\epsilon _{ijn}\;k_{i}^{\prime }\;k_{j};  \notag \\
R_{\beta }(m_{0},m_{i}) &=&m_{0}\;I+m_{i}\beta ^{i}=\left\vert
\begin{array}{rrrr}
m_{0} & m_{1} & m_{2} & m_{3} \\
-m_{1} & m_{0} & m_{3} & -m_{2} \\
-m_{2} & -m_{3} & m_{0} & m_{1} \\
-m_{3} & m_{2} & -m_{1} & m_{0}%
\end{array}%
\right\vert ,  \notag \\
m_{0}^{\prime \prime } &=&m_{0}^{\prime }\;m_{0}-m_{i}^{\prime
}\;m_{i},\qquad m_{n}^{\prime \prime }=m_{0}^{\prime }\;m_{n}+m_{n}^{\prime
}\;m_{0}-\epsilon _{ijn}\;m_{i}^{\prime }\;m_{j}.  \label{2.3}
\end{eqnarray}

Because the matrices of these two subgroups commute one with each other, we
can multiply them and, in such a way, we can obtain a new subgroup.
Moreover, this new subgroup allows us to impose the following constraints to
the parameters:%
\begin{equation*}
m_{0}=k_{0}^{\ast },\qquad m_{l}=-k_{l}^{\ast };
\end{equation*}%
These constraints lead to
\begin{eqnarray}
R(k,\bar{k}^{\ast }) &=&R_{\alpha }(k)\;R_{\beta }(\bar{k}^{\ast
})=(k_{0}\;I+k_{i}\alpha ^{i})\;(m_{0}\;I-k_{j}^{\ast }\beta ^{j})=  \notag
\\
&&k_{0}k_{0}^{\ast }+k_{0}^{\ast }k_{i}\;\;\alpha ^{i}-k_{0}k_{i}^{\ast
}\;\;\beta ^{i}-k_{i}k_{j}^{\ast }\;\;\alpha ^{i}\beta ^{j};  \label{2.5}
\end{eqnarray}%
\begin{equation*}
\begin{array}{l}
k_{0}k_{0}^{\ast }+k_{0}^{\ast }k_{i}\;\alpha ^{i}-k_{0}k_{i}^{\ast }\;\beta
^{i}=\medskip \\
=\left\vert
\begin{array}{cccc}
k_{0}k_{0}^{\ast } & k_{0}^{\ast }k_{1}-k_{0}k_{1}^{\ast } & k_{0}^{\ast
}k_{2}-k_{0}k_{2}^{\ast } & k_{0}^{\ast }k_{3}-k_{0}k_{3}^{\ast } \\
-k_{0}^{\ast }k_{1}+k_{0}k_{1}^{\ast } & k_{0}k_{0}^{\ast } & -k_{0}^{\ast
}k_{3}-k_{0}k_{3}^{\ast } & +k_{0}^{\ast }k_{2}+k_{0}k_{2}^{\ast } \\
-k_{0}^{\ast }k_{2}+k_{0}k_{2}^{\ast } & k_{0}^{\ast }k_{3}+k_{0}k_{3}^{\ast
} & k_{0}k_{0}^{\ast } & -k_{0}^{\ast }k_{1}-k_{0}k_{1}^{\ast } \\
-k_{0}^{\ast }k_{3}+k_{0}k_{3}^{\ast } & -k_{0}^{\ast
}k_{2}-k_{0}k_{2}^{\ast } & +k_{0}^{\ast }k_{1}+k_{0}k_{1}^{\ast } &
k_{0}k_{0}^{\ast }%
\end{array}%
\right\vert ,\medskip \\
-k_{i}k_{j}^{\ast }\;\;\alpha ^{i}\beta ^{j}=\medskip \\
=\left\vert
\begin{array}{rrrr}
k_{j}k_{j}^{\ast } & k_{2}k_{3}^{\ast }-k_{3}k_{2}^{\ast } &
k_{3}k_{1}^{\ast }-k_{1}k_{3}^{\ast } & k_{1}k_{2}^{\ast }-k_{2}k_{1}^{\ast }
\\
k_{2}k_{3}^{\ast }-k_{3}k_{2}^{\ast } & (k_{1}k_{1}^{\ast }-k_{2}k_{2}^{\ast
}-k_{3}k_{3}^{\ast }) & k_{1}k_{2}^{\ast }+k_{2}k_{1}^{\ast } &
k_{1}k_{3}^{\ast }+k_{3}k_{1}^{\ast } \\
k_{3}k_{1}^{\ast }-k_{1}k_{3}^{\ast } & k_{1}k_{2}^{\ast }+k_{2}k_{1}^{\ast }
& (k_{2}k_{2}^{\ast }-k_{1}k_{1}^{\ast }-k_{3}k_{3}^{\ast }) &
k_{2}k_{3}^{\ast }+k_{3}k_{2}^{\ast } \\
k_{1}k_{2}^{\ast }-k_{2}k_{1}^{\ast } & k_{1}k_{3}^{\ast }+k_{3}k_{1}^{\ast }
& k_{2}k_{3}^{\ast }+k_{3}k_{2}^{\ast } & (k_{3}k_{3}^{\ast
}-k_{1}k_{1}^{\ast }-k_{2}k_{2}^{\ast })%
\end{array}%
\right\vert .%
\end{array}%
\end{equation*}%
The matrices $R(k,\bar{k}^{\ast })$ determine transformations in the
4-dimensional space with one real and three imaginary coordinates.
Transition to all four real coordinates is achieved via
\begin{eqnarray}
y_{a}^{\prime } &=&R_{ab}\;y_{b},\qquad \left\vert
\begin{array}{c}
y_{0} \\
y_{1} \\
y_{2} \\
y_{3}%
\end{array}%
\right\vert =\left\vert
\begin{array}{cccc}
1 & 0 & 0 & 0 \\
0 & i & 0 & 0 \\
0 & 0 & i & 0 \\
0 & 0 & 0 & i%
\end{array}%
\right\vert \left\vert
\begin{array}{c}
x_{0} \\
x_{1} \\
x_{2} \\
x_{3}%
\end{array}%
\right\vert ,  \notag \\
Y &=&\Pi \;X,\qquad L(k,\bar{k}^{\ast })=\Pi ^{-1}R(k,\bar{k}^{\ast })\Pi ,
\label{2.6}
\end{eqnarray}%
\begin{equation*}
\begin{array}{l}
\Pi ^{-1}\;(k_{0}k_{0}^{\ast }+k_{0}^{\ast }k_{i}\;\;\alpha
^{i}-k_{0}k_{i}^{\ast }\;\;\beta ^{i})\;\Pi =\medskip \\
=\left\vert
\begin{array}{cccc}
k_{0}k_{0}^{\ast } & i(k_{0}^{\ast }k_{1}-k_{0}k_{1}^{\ast }) &
i(k_{0}^{\ast }k_{2}-k_{0}k_{2}^{\ast }) & i(k_{0}^{\ast
}k_{3}-k_{0}k_{3}^{\ast }) \\
i(k_{0}^{\ast }k_{1}-k_{0}k_{1}^{\ast }) & k_{0}k_{0}^{\ast } & -k_{0}^{\ast
}k_{3}-k_{0}k_{3}^{\ast } & +k_{0}^{\ast }k_{2}+k_{0}k_{2}^{\ast } \\
i(k_{0}^{\ast }k_{2}-k_{0}k_{2}^{\ast }) & k_{0}^{\ast
}k_{3}+k_{0}k_{3}^{\ast } & k_{0}k_{0}^{\ast } & -k_{0}^{\ast
}k_{1}-k_{0}k_{1}^{\ast } \\
i(k_{0}^{\ast }k_{3}-k_{0}k_{3}^{\ast }) & -k_{0}^{\ast
}k_{2}-k_{0}k_{2}^{\ast } & +k_{0}^{\ast }k_{1}+k_{0}k_{1}^{\ast } &
k_{0}k_{0}^{\ast }%
\end{array}%
\right\vert ,\medskip \\
\Pi ^{-1}\;(-k_{i}k_{j}^{\ast }\;\;\alpha ^{i}\beta ^{j})\Pi =\medskip \\
=\left\vert
\begin{array}{rrrr}
k_{j}k_{j}^{\ast } & i(+k_{2}k_{3}^{\ast }-k_{3}k_{2}^{\ast }) &
i(-k_{1}k_{3}^{\ast }+k_{3}k_{1}^{\ast }) & i(-k_{1}k_{2}^{\ast
}-k_{2}k_{1}^{\ast }) \\
-i(+k_{2}k_{3}^{\ast }-k_{3}k_{2}^{\ast }) & k_{1}k_{1}^{\ast
}-k_{2}k_{2}^{\ast }-k_{3}k_{3}^{\ast } & k_{1}k_{2}^{\ast
}+k_{2}k_{1}^{\ast } & k_{1}k_{3}^{\ast }+k_{3}k_{1}^{\ast } \\
-i(-k_{1}k_{3}^{\ast }+k_{3}k_{1}^{\ast }) & k_{1}k_{2}^{\ast
}+k_{2}k_{1}^{\ast } & k_{2}k_{2}^{\ast }-k_{1}k_{1}^{\ast
}-k_{3}k_{3}^{\ast } & k_{2}k_{3}^{\ast }+k_{3}k_{2}^{\ast } \\
-i(+k_{1}k_{2}^{\ast }-k_{2}k_{1}^{\ast }) & +k_{1}k_{3}^{\ast
}+k_{3}k_{1}^{\ast } & +k_{2}k_{3}^{\ast }+k_{3}k_{2}^{\ast } &
k_{3}k_{3}^{\ast }-k_{1}k_{1}^{\ast }-k_{2}k_{2}^{\ast }%
\end{array}%
\right\vert .%
\end{array}%
\end{equation*}%
We further specify a particular case:

\medskip \underline{$k_{0}\neq 0\;,\qquad k_{3}\neq 0$}
\begin{equation*}
L(k,\bar{k}^{\ast })=\left\vert
\begin{array}{cccc}
k_{0}k_{0}^{\ast }+k_{3}k_{3}^{\ast } & 0 & 0 & i(k_{0}^{\ast
}k_{3}-k_{0}k_{3}^{\ast }) \\
0 & k_{0}k_{0}^{\ast }-k_{3}k_{3}^{\ast } & -k_{0}^{\ast
}k_{3}-k_{0}k_{3}^{\ast } & 0 \\
0 & k_{0}^{\ast }k_{3}+k_{0}k_{3}^{\ast } & k_{0}k_{0}^{\ast
}-k_{3}k_{3}^{\ast } & 0 \\
i(k_{0}^{\ast }k_{3}-k_{0}k_{3}^{\ast }) & 0 & 0 & k_{0}k_{0}^{\ast
}+k_{3}k_{3}^{\ast }%
\end{array}%
\right\vert .
\end{equation*}%
For real parameters we get Euclidean rotations:\footnote{%
Note that for the need of polarization optics the transformations having
non-unit determinant may also be of interest.}%
\begin{eqnarray}
k_{0}^{\ast } &=&k_{0}=D\cos {\frac{\phi }{2}},\quad k_{3}^{\ast
}=k_{3}=D\sin {\frac{\phi }{2}},  \notag \\
L(k,\bar{k}^{\ast }) &=&D^{2}\left\vert
\begin{array}{cccc}
1 & 0 & 0 & 0 \\
0 & \cos \phi & -\sin \phi & 0 \\
0 & \sin \phi & \cos \phi & 0 \\
0 & 0 & 0 & 1%
\end{array}%
\right\vert ;  \label{2.9}
\end{eqnarray}
For complex parameters we get pseudo-Euclidean transformations:%
\begin{eqnarray}
k_{0} &=&k_{0}^{\ast }=D\;\cosh \;{\frac{\beta }{2}},\quad
k_{3}=-k_{3}^{\ast }=iD\;\sinh \;{\frac{\beta }{2}},  \notag \\
L &=&\Pi ^{-1}R\Pi =D^{2}\left\vert
\begin{array}{cccc}
\cosh \;\beta & 0 & 0 & -\;\sinh \;\beta \\
0 & 1 & 0 & 0 \\
0 & 0 & 1 & 0 \\
-\;\sinh \;\beta & 0 & 0 & \cosh \;\beta%
\end{array}%
\right\vert .  \label{2.10}
\end{eqnarray}%
The quantity $D$ determines the determinant of $L$, via
\begin{equation*}
\mbox{det}\;L=D^{8}.
\end{equation*}%
Let us consider now another particular example:

\medskip \underline{$k_{0}\neq 0\;,\qquad k_{1}\neq 0$}
\begin{equation*}
L=\Pi ^{-1}\;R\;\Pi =\left\vert
\begin{array}{cccc}
k_{0}k_{0}^{\ast }+k_{1}k_{1}^{\ast } & i(k_{0}^{\ast
}k_{1}-k_{0}k_{1}^{\ast }) & 0 & 0 \\
i(k_{0}^{\ast }k_{1}-k_{0}k_{1}^{\ast }) & k_{0}k_{0}^{\ast
}+k_{1}k_{1}^{\ast } & 0 & 0 \\
0 & 0 & k_{0}k_{0}^{\ast }-k_{1}k_{1}^{\ast } & -k_{0}^{\ast
}k_{1}-k_{0}k_{1}^{\ast } \\
0 & 0 & +k_{0}^{\ast }k_{1}+k_{0}k_{1}^{\ast } & k_{0}k_{0}^{\ast
}-k_{1}k_{1}^{\ast }%
\end{array}%
\right\vert ,
\end{equation*}%
whence it follows
\begin{eqnarray}
k_{0}^{\ast } &=&k_{0}=D\cos {\frac{\phi }{2}},\quad k_{1}^{\ast
}=k_{3}=D\sin {\frac{\phi }{2}},  \notag \\
L &=&D^{2}\left\vert
\begin{array}{cccc}
1 & 0 & 0 & 0 \\
0 & 1 & 0 & 0 \\
0 & 0 & \cos \phi & -\sin \phi \\
0 & 0 & \sin \phi & \cos \phi%
\end{array}%
\right\vert ;  \label{2.13} \\
k_{0} &=&k_{0}^{\ast }=D\;\cosh \;{\frac{\beta }{2}},\quad
k_{1}=-k_{1}^{\ast }=iD\;\sinh \;{\frac{\beta }{2}},  \notag \\
L &=&D^{2}\left\vert
\begin{array}{cccc}
\cosh \;\beta & -\;\sinh \;\beta & 0 & 0 \\
-\;\sinh \;\beta & \cosh \;\beta & 0 & 0 \\
0 & 0 & 1 & 0 \\
0 & 0 & 0 & 1%
\end{array}%
\right\vert .  \label{2.14}
\end{eqnarray}%
Let us write below the explicit form of the factors $R_{\alpha
}(k_{0},k_{j}) $ and $R_{\beta }(k_{0}^{\ast },-k_{j}^{\ast })$, after
performing the above similarity tramsformation:%
\begin{eqnarray}
K &=&\Pi ^{-1}\;R_{\alpha }(k_{0},k_{j})\Pi =\left\vert
\begin{array}{rrrr}
k_{0} & ik_{1} & ik_{2} & ik_{3} \\
ik_{1} & k_{0} & -k_{3} & k_{2} \\
ik_{2} & k_{3} & k_{0} & -k_{1} \\
ik_{3} & -k_{2} & k_{1} & k_{0}%
\end{array}%
\right\vert ,  \notag \\
K^{\ast } &=&\Pi ^{-1}\;R_{\beta }(k_{0}^{\ast },-k_{j}^{\ast })\Pi
=\left\vert
\begin{array}{rrrr}
k_{0}^{\ast } & -ik_{1}^{\ast } & -ik_{2}^{\ast } & -ik_{3}^{\ast } \\
-ik_{1}^{\ast } & k_{0}^{\ast } & -k_{3}^{\ast } & k_{2}^{\ast } \\
-ik_{2}^{\ast } & k_{3}^{\ast } & k_{0}^{\ast } & -k_{1}^{\ast } \\
-ik_{3}^{\ast } & -k_{2}^{\ast } & k_{1}^{\ast } & k_{0}^{\ast }%
\end{array}%
\right\vert ,  \label{2.15}
\end{eqnarray}

According to the afore mentioned considerations, any arbitrary
transformation of the Lorentz group may be factorized into two mutually
commuting and conjugate terms as%
\begin{equation*}
L=K\;K^{\ast }=K^{\ast }\;K.
\end{equation*}%
In Physics literature, that fact was firstly noted by Einstein and Mayer (in
1932--1933), while constructing the theory of semi-spinors \cite%
{1932-Einstein(1),1933-Einstein(1),1933-Einstein(2)}; further, a systematic
approach of the Lorentz group was developed by Fedorov \cite{4}.

\smallskip Let us write below the explicit form of the 2-parametric
commuting factors:%
\begin{eqnarray}
K_{1} &=&\left\vert
\begin{array}{rrrr}
k_{0} & ik_{1} & 0 & 0 \\
ik_{1} & k_{0} & 0 & 0 \\
0 & 0 & k_{0} & -k_{1} \\
0 & 0 & k_{1} & k_{0}%
\end{array}%
\right\vert ,\qquad K_{1}^{\ast }=\left\vert
\begin{array}{rrrr}
k_{0}^{\ast } & -ik_{1}^{\ast } & 0 & 0 \\
-ik_{1}^{\ast } & k_{0}^{\ast } & 0 & 0 \\
0 & 0 & k_{0}^{\ast } & -k_{1}^{\ast } \\
0 & 0 & k_{1}^{\ast } & k_{0}^{\ast }%
\end{array}%
\right\vert ,  \notag \\
K_{2} &=&\left\vert
\begin{array}{rrrr}
k_{0} & 0 & ik_{2} & 0 \\
0 & k_{0} & 0 & k_{2} \\
ik_{2} & 0 & k_{0} & 0 \\
0 & -k_{2} & 0 & k_{0}%
\end{array}%
\right\vert ,\qquad K^{\ast }=\left\vert
\begin{array}{rrrr}
k_{0}^{\ast } & 0 & -ik_{2}^{\ast } & 0 \\
0 & k_{0}^{\ast } & 0 & k_{2}^{\ast } \\
-ik_{2}^{\ast } & 0 & k_{0}^{\ast } & 0 \\
0 & -k_{2}^{\ast } & 0 & k_{0}^{\ast }%
\end{array}%
\right\vert ,  \notag \\
K_{3} &=&\left\vert
\begin{array}{rrrr}
k_{0} & 0 & 0 & ik_{3} \\
0 & k_{0} & -k_{3} & 0 \\
0 & k_{3} & k_{0} & 0 \\
ik_{3} & 0 & 0 & k_{0}%
\end{array}%
\right\vert ,\qquad K^{\ast }=\left\vert
\begin{array}{rrrr}
k_{0}^{\ast } & 0 & 0 & -ik_{3}^{\ast } \\
0 & k_{0}^{\ast } & -k_{3}^{\ast } & 0 \\
0 & k_{3}^{\ast } & k_{0}^{\ast } & 0 \\
-ik_{3}^{\ast } & 0 & 0 & k_{0}^{\ast }%
\end{array}%
\right\vert .  \label{2.17}
\end{eqnarray}

%=========================================================

\section{The action of the pseudo-Euclidean rotations on the partially\\ polarized light}

Let us consider the action of pure Lorentz transformations on 4-vectors in
the context of Stokes formalism, and firstly let us restrict to partially
polarized light. The corresponding Stokes 4-vectors are analogous to the
velocity 4-vectors of massive particles:%
\begin{equation*}
S^{a}=(I,I\mathbf{p}),\qquad U^{a}=(U^{0},U^{0}\mathbf{V});
\end{equation*}%
Moreover, these 4-vectors behave in the same way with respect to Lorentzian
transformations
\begin{equation*}
L=\left\vert
\begin{array}{cc}
\cosh \;\beta & -\mathbf{e}\;\sinh \;\beta \\
-\mathbf{e}\;\sinh \;\beta & [\delta _{ij}+(\cosh \;\beta -1)e_{i}e_{j}]\;%
\end{array}%
\right\vert ,\qquad \mathbf{e}^{2}=1.
\end{equation*}

They act on the Stokes vectors as follows:%
\begin{eqnarray}
I^{\prime } &=&I\;(\cosh \;\beta \;-\sinh \;\beta \;\mathbf{e}\mathbf{p}),
\notag \\
I^{\prime }\mathbf{p}^{\prime } &=&I\;[\;-\sinh \;\beta \;\mathbf{e}+\mathbf{%
p}+(\cosh \;\beta -1)\;\mathbf{e}(\mathbf{e}\mathbf{p})\;],  \notag
\end{eqnarray}

or
\begin{eqnarray}
I^{\prime } &=&I\;(\cosh \;\beta \;-\sinh \;\beta \;\mathbf{e}\mathbf{p}),
\notag \\
\mathbf{p}^{\prime } &=&{\frac{-\sinh \;\beta \;\mathbf{e}+\mathbf{p}+(\cosh
\;\beta -1)\;\mathbf{e}(\mathbf{e}\mathbf{p})}{\cosh \;\beta \;-\sinh
\;\beta \;\mathbf{e}\mathbf{p}}}.  \label{3.2}
\end{eqnarray}

Let us specify them in the following several special cases:

\medskip \underline{1}:
\begin{equation*}
\mathbf{e}\;\mathbf{p}=0,\qquad I^{\prime }=I\;\cosh \;\beta ,\qquad \mathbf{%
p}^{\prime }={\frac{-\sinh \;\beta \;\mathbf{e}+\mathbf{p}}{\cosh \;\beta }},
\end{equation*}%
where the intensity of the light beam increases and the polarization degree
decreases.

\medskip \underline{2}:
\begin{equation*}
\mathbf{p}=+p\;\mathbf{e},\qquad I^{\prime }=I\;(\cosh \;\beta \;-\sinh
\;\beta \;p),
\end{equation*}%
where the polarization vector changes according to the law
\begin{equation*}
\mathbf{p}^{\prime }=p^{\prime }\;\mathbf{e}={\frac{-\sinh \;\beta +\cosh
\;\beta \;p}{\cosh \;\beta \;-\sinh \;\beta \;p}}\;\mathbf{e}.
\end{equation*}%
Therefore the polarization degree varies as
\begin{equation*}
p^{\prime }=p{\frac{1-p^{-1}\tanh \;\beta }{1-\tanh\;\beta \;p}}.
\end{equation*}%
Note that when $\beta >0$, this decreases, and when $\beta <0$, then it
increases. In particular, there exists the \textit{"rest reference frame"}
where the partially polarized light becomes natural light:
\begin{eqnarray}
L_{0} &=&L_{0}(\beta _{0},\mathbf{n}),\qquad \tanh \;\beta _{0}=p,  \notag
\\
p^{\prime } &=&0,\qquad I^{\prime }=I\;(\cosh \;\beta _{0}\;-\sinh
_{0}\;\beta \;\tanh \;\beta _{0})={\frac{I}{\cosh \;\beta _{0}}},
\label{3.4c}
\end{eqnarray}%
and here the polarization degree equals to zero.

\medskip \underline{3}:
\begin{eqnarray}
\underline{\mathbf{p}=-p\;\mathbf{e}},\qquad I^{\prime } &=&I\;(\cosh
\;\beta \;+\sinh \;\beta \;p),  \notag \\
\mathbf{p}^{\prime } &=&p^{\prime }\;\mathbf{e}=-{\frac{\sinh \;\beta +\cosh
\;\beta \;p}{\cosh \;\beta \;+\sinh \;\beta \;p}}\;\mathbf{e}.  \label{3.5a}
\end{eqnarray}%
The polarization degree varies according to the law
\begin{equation*}
p^{\prime }=p{\frac{1+p^{-1}\tanh \;\beta }{1+\tanh \;\beta \;p}},
\end{equation*}%
where for $\beta >0$ increases, while for $\beta <0$ it decreases. Again,
there exist an analogue of the \textit{"rest reference frame" }$L_{0}$, in
which the polarization degree equals to zero.

\medskip Let us further consider the role of the relativistic ellipsoid in
optics. We start with the simplest case, $\mathbf{e}=(0,0,1)$, while the
formulas (3.2) give
\begin{eqnarray}
I^{\prime } &=&I\;(\cosh \;\beta \;-p_{3}\;\sinh \;\beta ),\qquad
p_{3}^{\prime }={\frac{\cosh \;\beta p_{3}-\sinh \;\beta }{\cosh \;\beta
-p_{3}\;\sinh \;\beta }},  \label{3.6} \\
p_{1}^{\prime } &=&{\frac{p_{1}}{(\cosh \;\beta \;-p_{3}\;\sinh \;\beta )}}%
,\qquad p_{2}^{\prime }={\frac{p_{2}}{(\cosh \;\beta \;-p_{3}\;\sinh \;\beta
)}}.  \label{3.7}
\end{eqnarray}%
Due to the basic property of the Lorentz matrices, the following identity
holds:%
\begin{equation*}
I^{^{\prime }2}(1-p^{^{\prime }2})=I^{2}(1-p^{2})\qquad \mbox{or}\qquad
1-p^{^{\prime }2}={\frac{1-p^{2}}{(\cosh \;\beta \;-p_{3}\;\sinh \;\beta
)^{2}}}.
\end{equation*}%
Therefore, the polarization degree transforms according to
\begin{equation}
p^{^{\prime }2}=1-{\frac{1-p^{2}}{(\cosh \;\beta \;-p_{3}\;\sinh \;\beta
)^{2}}}.  \label{3.8}
\end{equation}%
By expressing $p_{3}$ through $p_{3}^{\prime }$, we infer%
\begin{equation*}
p_{3}={\frac{\cosh \;\beta \;p_{3}^{\prime }+\sinh \;\beta }{\cosh \;\beta
+\sinh \;\beta \;p_{3}^{\prime }}}\qquad \Longrightarrow \qquad \cosh
\;\beta \;-p_{3}\;\sinh \;\beta ={\frac{1}{\cosh \;\beta +\sinh \;\beta
\;p_{3}^{\prime }}},
\end{equation*}%
and (\ref{3.8}) reduces to the form
\begin{equation}
p^{^{\prime }2}=1-(1-p^{2})(\cosh \;\beta +\sinh \;\beta \;p_{3}^{\prime
})^{2}.  \label{3.9}
\end{equation}%
Let us show that this equation (\ref{3.9}) describes an ellipsoid. Indeed, (%
\ref{3.9}) can be re-written as
\begin{equation*}
p_{1}^{^{\prime }2}+p_{2}^{^{\prime }2}+p_{3}^{^{\prime
}2}+(1-p^{2})2\;\cosh \;\beta \;\sinh \beta \;p_{3}^{\prime }+(1-p^{2})\sinh
^{2}\beta \;p_{3}^{^{\prime }2}=1-(1-p^{2})\cosh ^{2}\beta \;,
\end{equation*}%
or%
\begin{equation*}
\begin{array}{c}
p_{1}^{^{\prime }2}+p_{2}^{^{\prime }2}+(\cosh ^{2}\beta -p^{2}\;\sinh
^{2}\beta )\;\left[ p_{3}^{\prime }+{\dfrac{(1-p^{2})\sinh \;\beta \;\cosh
\;\beta }{\cosh ^{2}\beta -p^{2}\;\sinh ^{2}\beta }}\right] ^{2}=\medskip \\
=p^{2}\cosh ^{2}\beta -\sinh ^{2}\beta \;+{\dfrac{(1-p^{2})^{2}\sinh
^{2}\beta \;\cosh ^{2}\beta }{\cosh ^{2}\beta -p^{2}\;\sinh ^{2}\beta }},%
\end{array}%
\end{equation*}
and finally we obtain the ellipsoid equation%
\begin{equation}
p_{1}^{^{\prime }2}+p_{2}^{^{\prime }2}+(\cosh ^{2}\beta -p^{2}\;\sinh
^{2}\beta )\;(p_{3}^{\prime }+\gamma )^{2}={\frac{p^{2}}{\cosh ^{2}\beta
-p^{2}\sinh ^{2}\beta }},  \label{3.10}
\end{equation}%
where
\begin{equation*}
\gamma ={\frac{(1-p^{2})\;\sinh \;\beta \;\cosh \;\beta }{\cosh ^{2}\beta
-p^{2}\;\sinh ^{2}\beta }},\qquad \cosh ^{2}\beta -p^{2}\;\sinh ^{2}\beta
=\cosh ^{2}\beta (1-p^{2})+p^{2}>0.
\end{equation*}

Thus, the surface which has the form of a sphere ($\mathbf{p}^{2}=p^{2}$)
will transform under the action of a Mueller matrix of Lorentz type into
ellipsoid (\ref{3.10}).

\smallskip This result can be extended for more general Mueller matrices of
Lorentz type of arbitrary orientation:%
\begin{equation*}
I^{\prime }=I\;(\cosh \;\beta -\sinh \;\beta \;(\mathbf{e}\mathbf{p}%
)),\qquad \mathbf{p}^{\prime }={\frac{\mathbf{p}-\mathbf{e}\;\sinh \;\beta
+(\cosh \;\beta -1)\;\mathbf{e}\;(\mathbf{e}\mathbf{p})}{\cosh \;\beta
-\sinh \;\beta \;\mathbf{e}\;\mathbf{p}}}.
\end{equation*}

Considering the identity
\begin{equation}
I^{^{\prime }2}(1-p^{^{\prime }2})=I^{2}(1-p^{2})\qquad \Longrightarrow
\qquad 1-p^{^{\prime }2}={\frac{1-p^{2}}{[\cosh \;\beta -\sinh \;\beta \;(%
\mathbf{e}\mathbf{p})]^{2}}},  \label{3.11}
\end{equation}%
then, excluding the variable $\mathbf{p}$, we get:
\begin{equation*}
\mathbf{p}={\frac{\mathbf{p}^{\prime }+\mathbf{e}\;\sinh \;\beta +(\cosh
\;\beta -1)\;\mathbf{e}(\mathbf{e}\;\mathbf{p}^{\prime })}{\cosh \;\beta
+\sinh \;\beta \;\mathbf{e}\;\mathbf{p}^{\prime }}},
\end{equation*}%
or
\begin{equation*}
\cosh \;\beta -\sinh \;\beta (\mathbf{e}\mathbf{p})={\frac{1}{\cosh \;\beta
+\sinh \;\beta \;\mathbf{e}\mathbf{p}^{\prime }}}.
\end{equation*}%
From (\ref{3.11}) we get that
\begin{equation*}
1-p^{^{\prime }2}=(1-p^{2})\;(\cosh \;\beta +\sinh \;\beta \;\mathbf{e}%
\mathbf{p}^{\prime 2}.
\end{equation*}%
This describes an ellipsoid equation with the orientation governed by the
vector $\mathbf{e}$.

%=========================================================

\section{The action of a Lorentzian transformation on completely\\ polarized light}

Now let us consider the action of Lorentzian transformations on completely
polarized light (the analogous of the isotropic 4-vectors in Special
Relativity):%
\begin{equation*}
S^{a}=(I,I\mathbf{n}),\qquad \mathbf{n}^{2}=1.\eqno(4.1)
\end{equation*}%
With respect to transformations of kind%
\begin{equation*}
L=\left\vert
\begin{array}{cc}
\cosh \;\beta  & -\mathbf{e}\;\sinh \;\beta  \\
-\mathbf{e}\;\sinh \;\beta  & [\delta _{ij}+(\cosh \;\beta -1)e_{i}e_{j}]%
\end{array}%
\right\vert ,
\end{equation*}%
the Stokes 4-vector transforms as follows:%
\begin{equation*}
I^{\prime }=I\;(\cosh \;\beta \;-\sinh \;\beta \;\mathbf{e}\mathbf{n}%
),\qquad \mathbf{n}^{\prime }={\frac{-\sinh \;\beta \;\mathbf{e}+\mathbf{n}%
+(\cosh \;\beta -1)\;\mathbf{e}(\mathbf{e}\mathbf{n})}{\cosh \;\beta
\;-\sinh \;\beta \;\mathbf{e}\mathbf{n}}}.
\end{equation*}%
Let us specify now several special cases:\medskip

\noindent\underline{1}:
\begin{equation*}
\underline{\mathbf{e}\;\mathbf{n}=0},\qquad I^{\prime }=I\;\mbox{ ch}\;\beta
,\qquad \mathbf{n}^{\prime }={\frac{\mathbf{n}-\sinh \;\beta \;\mathbf{e}}{%
\cosh \;\beta }}.
\end{equation*}%
\underline{2}:
\begin{equation}
\underline{\mathbf{e}=+\;\mathbf{n}},\qquad I^{\prime -\beta },\qquad\mathbf{%
n}^{\prime }={\frac{-\sinh \;\beta \;\mathbf{n}+\mathbf{n}+(\cosh \;\beta
-1)\;\mathbf{n}}{\cosh \;\beta \;-\sinh \;\beta \;}}=+\mathbf{n}.
\label{4.4}
\end{equation}%
\underline{3}:
\begin{equation}
\underline{\mathbf{e}=-\;\mathbf{n}},\qquad I^{\prime +\beta },\qquad\mathbf{%
n}^{\prime }={\frac{\sinh \;\beta \;\mathbf{n}+\mathbf{n}+(\cosh \;\beta
-1)\;\mathbf{n}}{\cosh \;\beta \;+\sinh \;\beta \;}}=+\mathbf{n}  \label{4.5}
\end{equation}%
The known non-existence of the "rest reference frame" for massless particles
means, in the context of polarization optics, that one cannot completely
transform the polarized light into a natural one.

Note that under the action of Mueller transformations of Lorentz type, the
degree of the polarization of light does not change, that is $p^{\prime }=p=1
$.

%=========================================================

\section{On the deformations of the Stokes 4-vectors}

Now, let us consider in detail the Mueller matrices of diagonal form: --
these describe simple deformations of the Stokes 4-vectors. Firstly, let it
be the \underline{variant $U_{0}(i\lambda )$:}
\begin{equation*}
M=U_{0}(i\lambda )=\left\vert
\begin{array}{cccc}
e^{-\lambda } & 0 & 0 & 0 \\
0 & e^{-\lambda } & 0 & 0 \\
0 & 0 & e^{-\lambda } & 0 \\
0 & 0 & 0 & e^{-\lambda }%
\end{array}%
\right\vert ,\qquad S_{a}^{\prime }=e^{-\lambda }\;S_{a}.
\end{equation*}%
The inequalities (\ref{2.1a}) hold good:%
\begin{equation*}
S_{0}^{\prime }=e^{-\lambda }\;S_{0}\geq 0,\qquad S^{\prime 2}=e^{-2\lambda
}S_{0}^{2}\geq 0.
\end{equation*}

The action of such Mueller matrices on the light is given by the relations
\begin{equation*}
I^{\prime -\lambda }I,\qquad p_{i}^{\prime }=p_{i}.
\end{equation*}%
We further consider the \underline{variant $U_{2}^{B}(i\lambda )$}:
\begin{eqnarray}
M &=&U_{2}^{B}(i\lambda )=\left\vert
\begin{array}{cccc}
e^{\lambda } & 0 & 0 &  \\
0 & e^{-\lambda } & 0 & 0 \\
0 & 0 & e^{\lambda } & 0 \\
0 & 0 & 0 & e^{-\lambda }%
\end{array}%
\right\vert ,  \notag \\
S_{0}^{\prime } &=&e^{\lambda }\;S_{0},\qquad S_{1}^{\prime }=e^{-\lambda
}\;S_{1},\qquad S_{2}^{\prime }=e^{\lambda }\;S_{2},\qquad S_{3}^{\prime
}=e^{-\lambda }\;S_{3}.  \label{5.2a}
\end{eqnarray}

The restrictions (\ref{2.1a}) take the form (the first inequality always
holds good):
\begin{equation*}
S_{0}^{\prime }=e^{\lambda }\;S_{0}\geq 0,\qquad S^{\prime 2}=e^{2\lambda
}(S_{0}^{2}-S_{2}^{2})-e^{-2\lambda }\;(S_{1}^{2}+S_{3}^{2})\geq 0.
\end{equation*}%
The second one is equivalent to
\begin{equation*}
e^{4\lambda }\geq {\frac{S_{1}^{2}+S_{3}^{2}}{S_{0}^{2}-S_{2}^{2}}}.
\end{equation*}%
Since the initial light obeys the restriction
\begin{equation*}
S_{0}^{2}-S_{1}^{2}-S_{2}^{2}-S_{3}^{2}\geq 0\qquad \Longrightarrow \qquad
1\;\geq \;{\frac{S_{2}^{2}+S_{3}^{2}}{S_{0}^{2}-S_{1}^{2}}},
\end{equation*}%
where the inequality (2.2b) will be true when $\lambda $ is positive, that is%
\begin{equation*}
\lambda \in \lbrack 0,+\infty )
\end{equation*}%
The action of these Mueller matrices on the light is given by the relations
\begin{equation*}
I^{\prime \lambda }I,\qquad p_{1}^{\prime }=e^{-2\lambda }\;p_{1},\qquad
p_{2}^{\prime }=\;p_{2},\qquad p_{3}^{\prime }=e^{-2\lambda }\;p_{3}.
\end{equation*}%
Now, consider the \underline{ variant $U_{1}^{A}(i\lambda )$:}
\begin{eqnarray}
M &=&U_{1}^{A}(i\lambda )=\left\vert
\begin{array}{cccc}
e^{-\lambda } & 0 & 0 & 0 \\
0 & e^{-\lambda } & 0 & 0 \\
0 & 0 & e^{\lambda } & 0 \\
0 & 0 & 0 & e^{\lambda }%
\end{array}%
\right\vert ,  \notag \\
S_{0}^{\prime } &=&e^{-\lambda }\;S_{0},\qquad S_{1}^{\prime }=e^{-\lambda
}\;S_{1},\qquad S_{2}^{\prime }=e^{\lambda }\;S_{2},\qquad S_{3}^{\prime
}=e^{\lambda }\;S_{3},  \notag \\
S^{\prime 2} &=&e^{-2\lambda }(S_{0}^{2}-S_{1}^{2})-e^{2\lambda
}\;(S_{2}^{2}+S_{3}^{2})\geq 0.  \label{2.3a}
\end{eqnarray}%
The last inequality is equivalent to
\begin{equation*}
{\frac{S_{0}^{2}-S_{1}^{2}}{S_{2}^{2}+S_{3}^{2}}}\geq e^{4\lambda }\;\qquad
\Longrightarrow \qquad \lambda \in (-\infty ,0].
\end{equation*}%
The intensity of the light and the degree of polarization transform as
follows:%
\begin{equation*}
I^{\prime -\lambda }=I,\qquad \;p_{1}^{\prime }=p_{1},\qquad p_{2}^{\prime
}=e^{2\lambda }\;p_{2},\qquad p_{3}^{\prime }=e^{2\lambda }\ p_{3},
\end{equation*}

Now, consider the \underline{ variant $U_{3}^{C}(i\lambda )$}:%
\begin{eqnarray}
M &=&U_{3}^{C}(i\lambda )=\left\vert
\begin{array}{cccc}
e^{-\lambda } & 0 & 0 &  \\
0 & e^{\lambda } & 0 & 0 \\
0 & 0 & e^{\lambda } & 0 \\
0 & 0 & 0 & e^{-\lambda }%
\end{array}%
\right\vert .  \notag \\
S_{0}^{\prime } &=&e^{-\lambda }\;S_{0},\qquad S_{1}^{\prime }=e^{\lambda
}\;S_{1},\qquad S_{2}^{\prime }=e^{\lambda }\;S_{2},\qquad S_{3}^{\prime
}=e^{-\lambda }\;S_{3},  \notag \\
S^{\prime 2} &=&e^{-2\lambda }(S_{0}^{2}-S_{3}^{2})-e^{2\lambda
}\;(S_{1}^{2}+S_{2}^{2})\geq 0.  \label{5.4a}
\end{eqnarray}

The last inequality gives
\begin{equation*}
{\frac{S_{0}^{2}-S_{3}^{2}}{S_{1}^{2}+S_{2}^{2}}}\geq e^{4\lambda }\;\qquad
\Longrightarrow \qquad \lambda \in (-\infty ,0].
\end{equation*}%
The intensity of the light and the degree of polarization transform as
follows:%
\begin{equation*}
I^{\prime -\lambda }=I,\qquad p_{1}^{\prime }=e^{2\lambda }\;p_{1},\qquad
p_{2}^{\prime }=e^{2\lambda }\;p_{2},\qquad p_{3}^{\prime }=Ip_{3}\;.
\end{equation*}%
Instead of the four diagonal Mueller transformations described above, one
may introduce the following simpler four elementary deformations:%
\begin{eqnarray}
E_{0} &=&\left\vert
\begin{array}{cccc}
e^{\lambda } & 0 & 0 & 0 \\
0 & 1 & 0 & 0 \\
0 & 0 & 1 & 0 \\
0 & 0 & 0 & 1%
\end{array}%
\right\vert ,\qquad I^{\prime \lambda }=I,\;p_{1}^{\prime
}=p_{1},\;p_{2}^{\prime }=p_{2},\;p_{3}^{\prime }=p_{3},  \notag \\
E_{1} &=&\left\vert
\begin{array}{cccc}
1 & 0 & 0 & 0 \\
0 & e^{\lambda } & 0 & 0 \\
0 & 0 & 1 & 0 \\
0 & 0 & 0 & 1%
\end{array}%
\right\vert ,\qquad I^{\prime }=I,\;p_{1}^{\prime }=e^{\lambda
}p_{1},\;p_{2}^{\prime }=p_{2},\;p_{3}^{\prime }=p_{3},  \notag \\
E_{2} &=&\left\vert
\begin{array}{cccc}
e^{-\lambda } & 0 & 0 & 0 \\
0 & e^{\lambda } & 0 & 0 \\
0 & 0 & e^{\lambda } & 0 \\
0 & 0 & 0 & e^{-\lambda }%
\end{array}%
\right\vert ,\qquad I^{\prime }=I,\;p_{1}^{\prime }=p_{1},\;p_{2}^{\prime
}=e^{\lambda }p_{2},\;p_{3}^{\prime }=p_{3},  \notag \\
E_{3} &=&\left\vert
\begin{array}{cccc}
1 & 0 & 0 & 0 \\
0 & 1 & 0 & 0 \\
0 & 0 & 1 & 0 \\
0 & 0 & 0 & e^{\lambda }%
\end{array}%
\right\vert ,\qquad I^{\prime }=I,\;p_{1}^{\prime }=p_{1},\;p_{2}^{\prime
}=p_{2},\;p_{3}^{\prime }=e^{\lambda }p_{3}.  \label{5.5}
\end{eqnarray}

It is obvious that the above four matrices determine a 4-parametric Abelian
subgroup of the mutually commuting transformations.

%=========================================================

\section{On other subgroups of $SL(4,\mathbb{R})$}

Among the generators (1.5) one can emphasize the following triplets:%
\begin{eqnarray}
\mathbf{K} &=&\{A_{1}=\alpha _{1}\beta _{1},B_{2}=\alpha _{2}\beta
_{2},C_{3}=\alpha _{3}\beta _{3}\},  \notag \\
\mathbf{L} &=&\{C_{1}=\alpha _{1}\beta _{3},A_{2}=\alpha _{2}\beta
_{1},B_{3}=\alpha _{3}\beta _{2}\},  \notag \\
\mathbf{M} &=&\{B_{1}=\alpha _{1}\beta _{2},C_{2}=\alpha _{2}\beta
_{3},A_{3}=\alpha _{3}\beta _{1}\},  \label{6.1a}
\end{eqnarray}%
and
\begin{eqnarray}
\mathbf{K}^{\prime } &=&\{-C_{1}=-\alpha _{1}\beta _{3},-B_{2}=-\alpha
_{2}\beta _{2},-C_{3}=-\alpha _{3}\beta _{3}\},  \notag \\
\mathbf{L}^{\prime } &=&\{-B_{1}=-\alpha _{1}\beta _{2},-A_{2}=-\alpha
_{2}\beta _{1},-B_{3}=-\alpha _{3}\beta _{2}\},  \notag \\
\mathbf{M}^{\prime } &=&\{-A_{1}=-\alpha _{1}\beta _{1},-C_{2}=-\alpha
_{2}\beta _{3},-B_{3}=-\alpha _{3}\beta _{2}\}.  \label{6.1b}
\end{eqnarray}%
For all 3-vectors of the above generators, we can accordingly construct
multiplication laws of the same type:%
\begin{equation*}
\Gamma _{1}\Gamma _{2}=-\Gamma _{3},\qquad \Gamma _{2}\Gamma _{1}=-\Gamma
_{3},\qquad \Gamma _{1}\Gamma _{2}-\Gamma _{2}\Gamma _{1}=0,
\end{equation*}%
and similar ones provided by cyclic permutations. Within each triple, the
generators commute one with each other. This means that on the base of each
triplet of generators one can construct 1-, 2-, 3-parametric Abelian
subgroups.

\medskip Let us mention that using the 15 generators $\alpha _{i},\;\beta
_{i},\;A_{i},B_{i},C_{i}$, one can build 20 isomorphic to $SU(2)$ Lie
algebras:
\begin{eqnarray}
&&(\alpha _{1},\alpha _{2},\alpha _{3}),\qquad (\beta _{1},\beta _{2},\beta
_{3}),  \notag \\
&&(\alpha _{1},A_{2},A_{3}),\qquad (A_{1},\alpha _{2},A_{3}),\qquad
(A_{1},A_{2},\alpha _{3}),  \notag \\
&&(\alpha _{1},B_{2},B_{3}),\qquad (B_{1},\alpha _{2},B_{3}),\qquad
(B_{1},B_{2},\alpha _{3}),  \notag \\
&&(\alpha _{1},C_{2},C_{3}),\qquad (C_{1},\alpha _{2},C_{3}),\qquad
(C_{1},C_{2},\alpha _{3}),  \notag \\
&&(\beta _{1},B_{1},C_{1}),\qquad (\beta _{1},B_{2},C_{2}),\qquad (\beta
_{1},B_{3},C_{3}),  \notag \\
&&(A_{1},\beta _{2},C_{1}),\qquad (A_{2},\beta _{2},C_{2}),\qquad
(A_{3},\beta _{2},C_{3}),  \notag \\
&&(A_{1},B_{1},\beta _{3}),\qquad (A_{2},B_{2},\beta _{3}),\qquad
(A_{3},B_{3},\beta _{3}).  \label{6.2}
\end{eqnarray}%
All these say that the reserve of the possible subgroups of $SL(4,\mathbb{R})
$ is rather large, and that we may expect to separate some of them as being
appropriate to define Mueller-type subgroups.
%=========================================================

\section{Examination of 16 elementary one-parametric subgroups in $SL(4,%
\mathbb{R})$}

Let us consider the remained 12 one-parametric subgroups in $SL(4,\mathbb{R})
$:

\textbf{(7.1)}

\vspace{5mm}

\underline{Variant $U_{1}^{\alpha }(\phi )$:}
\begin{equation*}
M=U_{1}^{\alpha }(\phi )=\left\vert
\begin{array}{cccc}
\cos \phi  & \sin \phi  & 0 & 0 \\
-\sin \phi  & \cos \phi  & 0 & 0 \\
0 & 0 & \cos \phi  & -\sin \phi  \\
0 & 0 & \sin \phi  & \cos \phi
\end{array}%
\right\vert ,
\end{equation*}%
or with the use of the block form
\begin{eqnarray}
M &=&\left\vert
\begin{array}{cccc}
k_{0}+k_{3} & k_{1}+k_{2} & n_{0}+n_{3} & n_{1}+n_{2} \\
k_{1}-k_{2} & k_{0}-k_{3} & n_{1}-n_{2} & n_{0}-n_{3} \\
l_{0}+l_{3} & l_{1}+l_{2} & m_{0}+m_{3} & m_{1}+m_{2} \\
l_{1}-l_{2} & l_{0}-l_{3} & m_{1}-m_{2} & m_{0}-m_{3}%
\end{array}%
\right\vert ,  \notag \\
&&\left.
\begin{array}{rrrr}
k_{0}=\cos \phi \;, & \;k_{1}=0\;, & \;k_{2}=+\sin \phi \;, & \;k_{3}=0\;,
\\
m_{0}=\cos \phi \;, & \;m_{1}=0\;, & \;m_{2}=-\sin \phi \;, & \;m_{3}=0\;,
\\
n_{0}=0\;, & \;n_{1}=0\;, & \;n_{2}=0\;, & \;n_{3}=0\;, \\
l_{0}=0\;, & \;l_{1}=0\;, & \;l_{2}=0\;, & \;l_{3}=0\;.%
\end{array}%
\right.   \label{7.1.2}
\end{eqnarray}%
The restrictions (2.1) give
\begin{eqnarray}
S_{0} &\geq &0\;,\qquad S_{0}^{2}-S_{1}^{2}-S_{2}^{2}-S_{3}^{2}\geq 0\;,
\notag \\
S_{0}^{\prime } &=&\cos \phi S_{0}+\sin \phi S_{1}\geq 0\;,  \notag \\
(\cos \phi S_{0}+\sin \phi S_{1})^{2} &\geq &(-\sin \phi S_{0}+\cos \phi
S_{1})^{2}+S_{2}^{2}+S_{3}^{2}\;\;\Longrightarrow   \notag \\
\cos 2\phi (S_{0}^{2}-S_{1}^{2})+\sin 2\phi
\;2S_{0}S_{1}-S_{2}^{2}-S_{3}^{2} &\geq &0\;.  \notag
\end{eqnarray}%
With respect to the variables
\begin{equation*}
\tan \phi =x\;,\qquad p_{1}=a\;,\;p_{2}=b\;,\;p_{3}=c,
\end{equation*}%
the\ above inequalities take the form
\begin{eqnarray}
a\sin \phi +\cos \phi  &\geq &0\;,  \notag \\
{\frac{1-x^{2}}{1+x^{2}}}(1-a^{2})+{\frac{2x}{1+x^{2}}}\;2a-b^{2}-c^{2}
&\geq &0\;.  \label{7.1.3}
\end{eqnarray}%
The sets of solutions of the first inequality can be illustrated by the
following formulas and by Fig. 1 (the interval $\phi \in \lbrack 1,2\pi ]$
is divided into four parts):
\begin{eqnarray}
\underline{a>0},\qquad \cos \phi  &>&0,\;\phi \in \{I,IV\},\qquad x\geq -{%
\frac{1}{a}};  \notag \\
\underline{a>0},\qquad \cos \phi  &<&0,\;\phi \in \{II,III\},\qquad \ x\leq -%
{\frac{1}{a}};\;\;\;  \notag \\
\underline{a<0},\qquad \cos \phi  &>&0,\;\phi \in \{I,IV\},\qquad x\leq -{%
\frac{1}{a}};  \notag \\
\underline{a<0},\qquad \cos \phi  &<&0,\;\phi \in \{II,III\},\qquad x\geq -{%
\frac{1}{a}}.  \label{7.1.4}
\end{eqnarray}%
The quadratic inequality (\ref{7.1.3}) is equivalent to
\begin{equation*}
(a^{2}-1-b^{2}-c^{2})x^{2}+4ax+(1-a^{2}-b^{2}-c^{2}))\geq 0.
\end{equation*}%
whose solution is $x\in \lbrack x_{1},x_{2}]$, where $x_{1},$ $x_{2}$ are
the roots of the equation
\begin{equation*}
x^{2}-2x{\frac{2a}{b^{2}+c^{2}+1-a^{2}}}-{\frac{1-a^{2}-b^{2}-c^{2}}{%
b^{2}+c^{2}+1-a^{2}}}=0.
\end{equation*}%
These are the real numbers
\begin{eqnarray}
x_{1} &=&{\frac{2a-\sqrt{4a^{2}+(1-b^{2}-c^{2}-a^{2})\;(b^{2}+c^{2}+1-a^{2})}%
}{b^{2}+c^{2}+1-a^{2}}},  \notag \\
x_{2} &=&{\frac{2a+\sqrt{4a^{2}+(1-b^{2}-c^{2}-a^{2})\;(b^{2}+c^{2}+1-a^{2})}%
}{b^{2}+c^{2}+1-a^{2}}}.  \label{7.1.5}
\end{eqnarray}%
In the case of completely polarized light, the formulas become much simpler:
\begin{equation*}
b^{2}+c^{2}+a^{2}=1,\qquad x_{1,2}=\left\{ 0\;,\;{\frac{2a}{1-a^{2}}}%
\right\} ;
\end{equation*}%
and two different possibilities should be distinguished
\begin{eqnarray}
a &>&0,\qquad 0\leq x\leq {\frac{2a}{1-a^{2}}}\qquad \{\;I,III\;\};  \notag
\\
a &<&0,\qquad {\frac{2a}{1-a^{2}}}\leq x\leq 0\qquad \{\;II,IV\;\};  \notag
\\
a &=&0,\qquad x_{1}=x_{2}=0\qquad (\mbox{degenerate case}).  \notag
\end{eqnarray}

For $a>0$, the joint solution of both inequalities in (\ref{7.1.3}) are
possible only in the quadrant I:
\begin{equation*}
a>0,\qquad 0\leq x\leq {\frac{2a}{1-a^{2}}}.
\end{equation*}

For $a<0$, the joint solution of both inequalities in (\ref{7.1.3}) is
possible only in the quadrant IV:
\begin{equation*}
a<0,\qquad {\frac{2a}{1-a^{2}}}\leq x\leq 0.
\end{equation*}

Let us investigate now the more complicated case of partially polarized
light. Since for any value of $a$ the root $x_{1}$ is negative, and the root
$x_{2}$ is positive, turning to Fig. 1, we see that the joint solutions of
both inequalities in (\ref{7.1.3}) are possible.

\vspace{20mm}

\begin{figure}[tbp]
\unitlength=0.27mm%
\begin{picture}(160,60)(-20,-20)
\special{em:linewidth 0.4pt} \linethickness{0.4pt}
{\bf\color[rgb]{0.5,0.3,0.9}
\put(+230,+90){\underline{$a>0$}}   }
\put(-30,+90){$x=\tan \phi$}
\put(+230,-10){$\phi$}
{\bf\color[rgb]{0.2,0.8,0.2}
\put(0,-12){\circle*{3}}  \put(-35,-16){$-a^{-1}$}
\put(0,-12){\line(1,0){200}}  }
\put(0,0){\vector(1,0){220}} \put(0,-80){\vector(0,1){160}}
{\bf\color[rgb]{0.5,0.3,0.9}
\put(0,0){\circle*{2}}
\put(10,4){\circle*{2}}
\put(30,12){\circle*{2}}
\put(40,30){\circle*{2}}
\put(44,50){\circle*{2}}
\put(45,70){\circle*{2}}
}

\put(0,0){\circle*{2}}
\put(90,-4){\circle*{2}}

{\bf\color[rgb]{0.5,0.3,0.9}
\put(70,-12){\circle*{2}}
\put(60,-30){\circle*{2}}
\put(56,-50){\circle*{2}}
\put(55,-70){\circle*{2}}  }

\put(100,0){\circle*{2}}
\put(110,4){\circle*{2}}
\put(130,12){\circle*{2}}
\put(140,30){\circle*{2}}
\put(144,50){\circle*{2}}
\put(145,70){\circle*{2}}

{\bf\color[rgb]{0.5,0.3,0.9}
\put(100,0){\circle*{2}}
\put(190,-4){\circle*{2}}   }

\put(170,-12){\circle*{2}}
\put(160,-30){\circle*{2}}
\put(156,-50){\circle*{2}}
\put(155,-70){\circle*{2}}

 \put(50,-80){\line(0,1){160}}
 \put(100,-80){\line(0,1){160}}
 \put(150,-80){\line(0,1){160}}
 \put(200,-80){\line(0,1){160}}

{\bf\color[rgb]{0.5,0.3,0.9}

\put(0,-0.2){\line(1,0){70}}
\put(0,-0.4){\line(1,0){70}}
\put(0,-0.6){\line(1,0){70}}
\put(0,+0.2){\line(1,0){70}}
\put(0,+0.4){\line(1,0){70}}
\put(0,+0.6){\line(1,0){70}}

\put(170,-0.2){\line(1,0){30}}
\put(170,-0.4){\line(1,0){30}}
\put(170,-0.6){\line(1,0){30}}
\put(170,+0.2){\line(1,0){30}}
\put(170,+0.4){\line(1,0){30}}
\put(170,+0.6){\line(1,0){30}}   }

\put(+25,-100){$I$}  \put(+75,-100){$II$} \put(+125,-100){$III$} \put(+175,-100){$IV$}

{\bf\color[rgb]{0.9,0,0}
\put(+530,+90){\underline{$a<0$}}}

{\bf\color[rgb]{0.2,0.8,0.2}

\put(300,+10){\circle*{3}}  \put(265,+16){$-a^{-1}$}
\put(300,+10){\line(1,0){200}}  }

\put(+285,+90){$x=\tan \phi$}
\put(+530,-10){$\phi$}

\put(300,0){\vector(1,0){220}} \put(300,-80){\vector(0,1){160}}

{\bf\color[rgb]{0.9,0,0}
\put(300,0){\circle*{2}}
\put(310,4){\circle*{2}}
\put(330,12){\circle*{2}}  }

\put(340,30){\circle*{2}}
\put(344,50){\circle*{2}}
\put(345,70){\circle*{2}}

\put(390,0){\circle*{2}}
\put(390,-4){\circle*{2}}
\put(370,-12){\circle*{2}}
\put(360,-30){\circle*{2}}
\put(356,-50){\circle*{2}}
\put(355,-70){\circle*{2}}

\put(400,0){\circle*{2}}
\put(410,4){\circle*{2}}

{\bf\color[rgb]{0.9,0,0}

\put(430,12){\circle*{2}}
\put(440,30){\circle*{2}}
\put(444,50){\circle*{2}}
\put(445,70){\circle*{2}}  }

\put(400,0){\circle*{2}}

{\bf\color[rgb]{0.9,0,0}
\put(490,-4){\circle*{2}}
\put(470,-12){\circle*{2}}
\put(460,-30){\circle*{2}}
\put(456,-50){\circle*{2}}
\put(455,-70){\circle*{2}}}

 \put(350,-80){\line(0,1){160}}
 \put(400,-80){\line(0,1){160}}
 \put(450,-80){\line(0,1){160}}
 \put(500,-80){\line(0,1){160}}

{\bf\color[rgb]{0.9,0,0}

\put(300,-0.2){\line(1,0){30}}
\put(300,-0.4){\line(1,0){30}}
\put(300,-0.6){\line(1,0){30}}
\put(300,+0.4){\line(1,0){30}}
\put(300,+0.2){\line(1,0){30}}
\put(300,+0.6){\line(1,0){30}}

\put(430,-0.2){\line(1,0){70}}
\put(430,-0.4){\line(1,0){70}}
\put(430,-0.6){\line(1,0){70}}
\put(430,+0.2){\line(1,0){70}}
\put(430,+0.4){\line(1,0){70}}
\put(430,+0.6){\line(1,0){70}}  }

\put(+325,-100){$I$}  \put(+375,-100){$II$} \put(+425,-100){$III$} \put(+475,-100){$IV$}
\end{picture}
\par
\vspace{30mm}
\caption{Solutions of inequalities (7.1.4) }
\label{fig:1-1}
\end{figure}
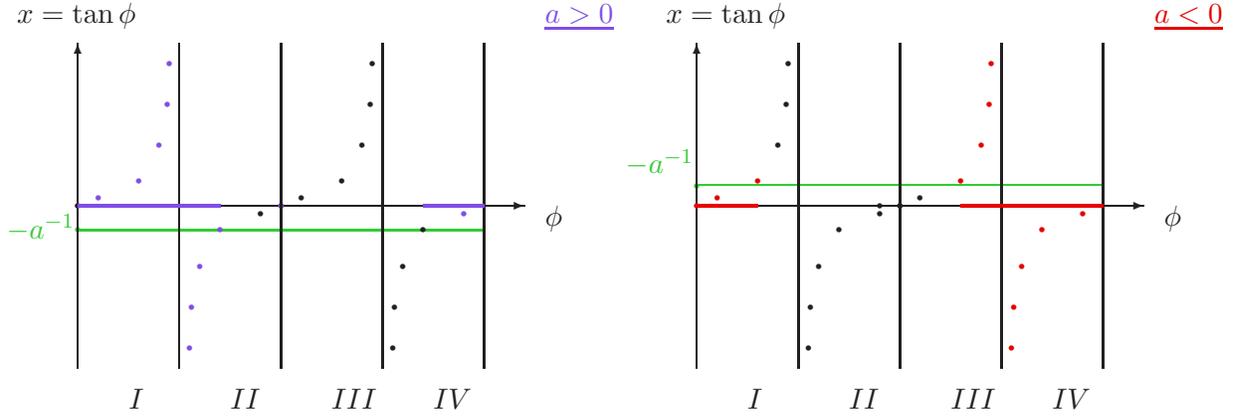

\vspace{5mm}

\textbf{(7.2)}

\vspace{5mm}

\underline{Variant $U_{2}^{\alpha }(\phi )$ }
\begin{equation*}
M=U_{2}^{\alpha }(-\phi )=\left\vert
\begin{array}{cccc}
\cos \phi  & 0 & \sin \phi  & 0 \\
0 & \cos \phi  & 0 & \sin \phi  \\
-\sin \phi  & 0 & \cos \phi  & 0 \\
0 & -\sin \phi  & 0 & \cos \phi
\end{array}%
\right\vert ,
\end{equation*}%
or in block form
\begin{eqnarray}
M &=&\left\vert
\begin{array}{cccc}
k_{0}+k_{3} & k_{1}+k_{2} & n_{0}+n_{3} & n_{1}+n_{2} \\
k_{1}-k_{2} & k_{0}-k_{3} & n_{1}-n_{2} & n_{0}-n_{3} \\
l_{0}+l_{3} & l_{1}+l_{2} & m_{0}+m_{3} & m_{1}+m_{2} \\
l_{1}-l_{2} & l_{0}-l_{3} & m_{1}-m_{2} & m_{0}-m_{3}%
\end{array}%
\right\vert ,  \notag \\
&&\left.
\begin{array}{rrrr}
k_{0}=\cos \phi , & \;k_{1}=0, & \;k_{2}=0, & \;k_{3}=0, \\
m_{0}=\cos \phi , & \;m_{1}=0, & \;m_{2}=0, & \;m_{3}=0, \\
n_{0}=\sin \phi , & \;n_{1}=0, & \;n_{2}=0, & \;n_{3}=0, \\
l_{0}=-\sin \phi , & \;l_{1}=0, & \;l_{2}=0, & \;l_{3}=0.%
\end{array}%
\right.   \label{7.2.2}
\end{eqnarray}%
The restrictions (\ref{2.1a}) lead to%
\begin{equation*}
\begin{array}{c}
S_{0}\geq 0\;,\qquad S_{0}^{2}-S_{1}^{2}-S_{2}^{2}-S_{3}^{2}\geq 0,\medskip
\\
S_{0}^{\prime ^{\prime }2}-S_{1}^{^{\prime }2}\geq S_{2}^{^{\prime
}2}+S_{3}^{^{\prime }2},\medskip \\
S_{0}^{\prime }=\cos \phi S_{0}+\sin \phi S_{2}\geq 0,\medskip \\
(\cos \phi S_{0}+\sin \phi S_{2})^{2}-(\cos \phi S_{1}+\sin \phi S_{3})^{2}-%
\medskip \\
(-\sin \phi S_{0}+\cos \phi S_{2})^{2}-(-\sin \phi S_{1}+\cos \phi
S_{3})^{2}>0\Longrightarrow \medskip \\
\Longrightarrow \cos 2\phi (S_{0}^{2}-S_{2}^{2})+\sin 2\phi
\;2S_{0}S_{2}-S_{1}^{2}-S_{3}^{2}\geq 0\;.%
\end{array}%
\end{equation*}

In the variables $a,b,c,x$, these inequalities take the form
\begin{eqnarray}
a^{2}+b^{2}+c^{2} &\leq &1,\qquad \cos \phi +b\;\sin \phi \geq 0,  \notag \\
{\frac{1-x^{2}}{1+x^{2}}}(1-b^{2})+{\frac{2x}{1+x^{2}}}\;2b-a^{2}-c^{2}
&\geq &0.  \label{7.2.3}
\end{eqnarray}%
They differ from the previous ones from (\ref{7.1.3}) only in formal
notation, and therefore the results will be much the same.

\vspace{5mm}

\textbf{(7.3)}

\vspace{5mm}

\underline{Variant $U_{3}^{\alpha }(\phi )$}
\begin{equation*}
U_{3}^{\alpha }(\phi )=\left\vert
\begin{array}{cccc}
\cos \phi  & 0 & 0 & \sin \phi  \\
0 & \cos \phi  & -\sin \phi  & 0 \\
0 & \sin \phi  & \cos \phi  & 0 \\
-\sin \phi  & 0 & 0 & \cos \phi
\end{array}%
\right\vert ,
\end{equation*}%
or in block form
\begin{eqnarray}
M &=&\left\vert
\begin{array}{cccc}
k_{0}+k_{3} & k_{1}+k_{2} & n_{0}+n_{3} & n_{1}+n_{2} \\
k_{1}-k_{2} & k_{0}-k_{3} & n_{1}-n_{2} & n_{0}-n_{3} \\
l_{0}+l_{3} & l_{1}+l_{2} & m_{0}+m_{3} & m_{1}+m_{2} \\
l_{1}-l_{2} & l_{0}-l_{3} & m_{1}-m_{2} & m_{0}-m_{3}%
\end{array}%
\right\vert ,  \notag \\
&&\left.
\begin{array}{rrrr}
k_{0}=\cos \phi , & \;k_{1}=0, & \;k_{2}=0, & \;k_{3}=0, \\
m_{0}=\cos \phi , & \;m_{1}=0, & \;m_{2}=0, & \;m_{3}=0, \\
n_{0}=0, & \;n_{1}=0, & \;n_{2}=\sin \phi , & \;n_{3}=0, \\
l_{0}=0, & \;l_{1}=0, & \;l_{2}=\sin \phi , & \;l_{3}=0.%
\end{array}%
\right.   \label{7.3.2}
\end{eqnarray}%
The restrictions (\ref{2.1a}) lead to
\begin{eqnarray}
S_{0} &\geq &0\;,\qquad S_{0}^{2}-S_{1}^{2}-S_{2}^{2}-S_{3}^{2}\geq 0\;,
\notag \\
S_{0}^{\prime } &=&\cos \phi S_{0}+\sin \phi S_{3}\geq 0\;,  \notag \\
(\cos \phi S_{0}+\sin \phi S_{3})^{2} &\geq &(-\sin \phi S_{0}+\cos \phi
S_{3})^{2}+S_{1}^{2}+S_{2}^{2}\Longrightarrow   \notag \\
\Longrightarrow \cos 2\phi (S_{0}^{2}-S_{3}^{2})+\sin 2\phi \;2S_{0}S_{3} &\geq
&S_{1}^{2}+S_{2}^{2}\;.  \notag
\end{eqnarray}%
In the variables $a,b,c,x$, the inequalities take the form
\begin{eqnarray}
\cos \phi +c\;\sin \phi  &\geq &,  \notag \\
{\frac{1-x^{2}}{1+x^{2}}}(1-c^{2})+{\frac{2x}{1+x^{2}}}\;2c-a^{2}-b^{2}
&\geq &0.  \label{7.3.3}
\end{eqnarray}%
These differ from the previous ones from (\ref{7.1.3}) only in formal
notation, and therefore the results will be much the same.

\vspace{5mm}

\textbf{(7.4)}

\vspace{5mm}

\underline{Variant $U_{1}^{\beta }(\phi )$}
\begin{equation*}
M=U_{1}^{\beta }(\phi )=\left\vert
\begin{array}{cccc}
\cos \phi  & \sin \phi  & 0 & 0 \\
-\sin \phi  & \cos \phi  & 0 & 0 \\
0 & 0 & \cos \phi  & \sin \phi  \\
0 & 0 & -\sin \phi  & \cos \phi
\end{array}%
\right\vert ,
\end{equation*}%
or in block form
\begin{eqnarray}
M &=&\left\vert
\begin{array}{cccc}
k_{0}+k_{3} & k_{1}+k_{2} & n_{0}+n_{3} & n_{1}+n_{2} \\
k_{1}-k_{2} & k_{0}-k_{3} & n_{1}-n_{2} & n_{0}-n_{3} \\
l_{0}+l_{3} & l_{1}+l_{2} & m_{0}+m_{3} & m_{1}+m_{2} \\
l_{1}-l_{2} & l_{0}-l_{3} & m_{1}-m_{2} & m_{0}-m_{3}%
\end{array}%
\right\vert ,  \notag \\
&&\left.
\begin{array}{rrrr}
k_{0}=\cos \phi , & \;k_{1}=0, & \;k_{2}=+\sin \phi , & \;k_{3}=0, \\
m_{0}=\cos \phi , & \;m_{1}=0, & \;m_{2}=+\sin \phi , & \;m_{3}=0, \\
n_{0}=0, & \;n_{1}=0, & \;n_{2}=0, & \;n_{3}=0, \\
l_{0}=0, & \;l_{1}=0, & \;l_{2}=0, & \;l_{3}=0.%
\end{array}%
\right.   \label{7.4.2}
\end{eqnarray}%
The restrictions (\ref{2.1a}) give
\begin{eqnarray}
S_{0} &\geq &0\;,\qquad S_{0}^{2}-S_{1}^{2}\geq S_{2}^{2}+S_{3}^{2}\;,
\notag \\
S_{0}^{\prime } &=&\cos \phi S_{0}+\sin \phi S_{1}\geq 0\;,  \notag \\
(\cos \phi S_{0}+\sin \phi S_{1})^{2} &\geq &(-\sin \phi S_{0}+\cos \phi
S_{1})^{2}+S_{2}^{2}+S_{3}^{2}\Longrightarrow   \notag \\
\cos 2\phi (S_{0}^{2}-S_{1}^{2})+\sin 2\phi \;2S_{0}S_{1} &\geq
&S_{2}^{2}+S_{3}^{2}\;.  \notag
\end{eqnarray}%
With respect to the variables $a,b,c,x$, the inequalities become
\begin{eqnarray}
\cos \phi +a\;\sin \phi  &\geq &0,  \notag \\
{\frac{1-x^{2}}{1+x^{2}}}(1-a^{2})+{\frac{2x}{1+x^{2}}}\;2a-b^{2}-c^{2}
&\geq &0.  \label{7.4.3}
\end{eqnarray}%
These differs from the previous ones from (\ref{7.1.3}) only in formal
notation, and therefore the results will be much the same.

\vspace{5mm}

\textbf{(7.5)}

\vspace{5mm}

\underline{Variant $U_{2}^{\beta }(\phi )$}
\begin{equation*}
M=U_{2}^{\beta }(\phi )=\left\vert
\begin{array}{cccc}
\cos \phi  & 0 & \sin \phi  & 0 \\
0 & \cos \phi  & 0 & -\sin \phi  \\
-\sin \phi  & 0 & \cos \phi  & 0 \\
0 & \sin \phi  & 0 & \cos \phi
\end{array}%
\right\vert ,
\end{equation*}%
or in block form
\begin{eqnarray}
M &=&\left\vert
\begin{array}{cccc}
k_{0}+k_{3} & k_{1}+k_{2} & n_{0}+n_{3} & n_{1}+n_{2} \\
k_{1}-k_{2} & k_{0}-k_{3} & n_{1}-n_{2} & n_{0}-n_{3} \\
l_{0}+l_{3} & l_{1}+l_{2} & m_{0}+m_{3} & m_{1}+m_{2} \\
l_{1}-l_{2} & l_{0}-l_{3} & m_{1}-m_{2} & m_{0}-m_{3}%
\end{array}%
\right\vert ,  \notag \\
&&\left.
\begin{array}{rrrr}
k_{0}=\cos \phi , & \;k_{1}=0, & \;k_{2}=0, & \;k_{3}=0, \\
m_{0}=\cos \phi , & \;m_{1}=0, & \;m_{2}=0, & \;m_{3}=0, \\
n_{0}=0, & \;n_{1}=0, & \;n_{2}=0, & \;n_{3}=\sin \phi , \\
l_{0}=0, & \;l_{1}=0, & \;l_{2}=0, & \;l_{3}=-\sin \phi .%
\end{array}%
\right.   \label{7.5.2}
\end{eqnarray}%
The restrictions (\ref{2.1a}) give
\begin{eqnarray}
S_{0} &\geq &0\;,\qquad S_{0}^{2}-S_{1}^{2}-S_{2}^{2}-S_{3}^{2}\geq 0,
\notag \\
S_{0}^{\prime ^{\prime }2}{}_{0}-S_{1}^{^{\prime }2} &\geq &S_{2}^{^{\prime
}2}+S_{3}^{^{\prime }2},  \notag \\
S_{0}^{\prime } &=&\cos \phi S_{0}+\sin \phi S_{2}\geq 0,  \notag \\
&&(\cos \phi S_{0}+\sin \phi S_{2})^{2}-(\cos \phi S_{1}-\sin \phi
S_{3})^{2}-  \notag \\
(-\sin \phi S_{0}+\cos \phi S_{2})^{2}-(\sin \phi S_{1}+\cos \phi S_{3})^{2}
&>&0\qquad \Longrightarrow   \notag \\
\Longrightarrow \cos 2\phi (S_{0}^{2}-S_{2}^{2})+\sin 2\phi
\;2S_{0}S_{2}-S_{1}^{2}-S_{3}^{2} &\geq &0.  \notag
\end{eqnarray}%
In the variables $a,b,c,x$ the above inequalities take the form
\begin{eqnarray}
a^{2}+b^{2}+c^{2} &\leq &1,\qquad \cos \phi -b\;\sin \phi \geq 0,  \notag \\
{\frac{1-x^{2}}{1+x^{2}}}(1-b^{2})+{\frac{2x}{1+x^{2}}}\;2b-a^{2}-c^{2}
&\geq &0.  \label{7.5.3}
\end{eqnarray}%
They differ from the previous ones from (\ref{7.1.3}) only in formal
notation, therefore the results will be much the same.

\vspace{5mm}

\textbf{(7.6)}

\vspace{5mm}

\underline{Variant $U_{3}^{\beta }(\phi )$}
\begin{equation*}
M=U_{3}^{\beta }(\phi )=\left\vert
\begin{array}{cccc}
\cos \phi  & 0 & 0 & \sin \phi  \\
0 & \cos \phi  & \sin \phi  & 0 \\
0 & -\sin \phi  & \cos \phi  & 0 \\
-\sin \phi  & 0 & 0 & \cos \phi
\end{array}%
\right\vert ,
\end{equation*}%
or in block form
\begin{eqnarray}
M &=&\left\vert
\begin{array}{cccc}
k_{0}+k_{3} & k_{1}+k_{2} & n_{0}+n_{3} & n_{1}+n_{2} \\
k_{1}-k_{2} & k_{0}-k_{3} & n_{1}-n_{2} & n_{0}-n_{3} \\
l_{0}+l_{3} & l_{1}+l_{2} & m_{0}+m_{3} & m_{1}+m_{2} \\
l_{1}-l_{2} & l_{0}-l_{3} & m_{1}-m_{2} & m_{0}-m_{3}%
\end{array}%
\right\vert ,  \notag \\
&&\left.
\begin{array}{rrrr}
k_{0}=\cos \phi , & \;k_{1}=0, & \;k_{2}=0, & \;k_{3}=0, \\
m_{0}=\cos \phi , & \;m_{1}=0, & \;m_{2}=0, & \;m_{3}=0, \\
n_{0}=0, & \;n_{1}=\sin \phi , & \;n_{2}=0, & \;n_{3}=0, \\
l_{0}=0, & \;l_{1}=-\sin \phi , & \;l_{2}=0, & \;l_{3}=0.%
\end{array}%
\right.   \label{7.6.2}
\end{eqnarray}%
The restrictions (\ref{2.1a}) give%
\begin{equation*}
\begin{array}{c}
S_{0}\geq 0,\qquad S_{0}^{2}-S_{1}^{2}-S_{2}^{2}-S_{3}^{2}\geq 0,\medskip \\
S_{0}^{\prime ^{\prime }2}{}_{0}-S_{1}^{^{\prime }2}\geq S_{2}^{^{\prime
}2}+S_{3}^{^{\prime }2},\medskip \\
S_{0}^{\prime }=\cos \phi S_{0}+\sin \phi S_{3}\geq 0,\medskip \\
(\cos \phi S_{0}+\sin \phi S_{3})^{2}-(\cos \phi S_{1}+\sin \phi S_{2})^{2}-%
\medskip \\
(-\sin \phi S_{1}+\cos \phi S_{2})^{2}-(-\sin \phi S_{0}+\cos \phi
S_{3})^{2}>0\Longrightarrow \medskip \\
\Longrightarrow \cos 2\phi (S_{0}^{2}-S_{3}^{2})+\sin 2\phi
\;2S_{0}S_{3}-S_{1}^{2}-S_{2}^{2}\geq 0.%
\end{array}%
\end{equation*}
In the variables $a,b,c,x$ the inequalities take the form
\begin{eqnarray}
a^{2}+b^{2}+c^{2} &\leq &1,\qquad \cos \phi +c\;\sin \phi \geq 0,  \notag \\
{\frac{1-x^{2}}{1+x^{2}}}(1-c^{2})+{\frac{2x}{1+x^{2}}}\;2c-a^{2}-b^{2}
&\geq &0.  \label{7.6.3}
\end{eqnarray}%
These differ from the previous ones from (\ref{7.1.3}) only in formal
notation, therefore the results will be much the same.

\vspace{5mm}

\textbf{(7.7)}

\vspace{5mm}

\underline{Variant $U_{2}^{A}(-i\beta )$ }
\begin{equation*}
U_{2}^{A}(i\beta )=\left\vert
\begin{array}{cccc}
\cosh \;\beta  & 0 & 0 & \sinh \;\beta  \\
0 & \cosh \;\beta  & -\sinh \;\beta  & 0 \\
0 & -\sinh \;\beta  & \cosh \;\beta  & 0 \\
\sinh \;\beta  & 0 & 0 & \cosh \;\beta
\end{array}%
\right\vert ,
\end{equation*}%
or in block form
\begin{eqnarray}
M &=&\left\vert
\begin{array}{cccc}
k_{0}+k_{3} & k_{1}+k_{2} & n_{0}+n_{3} & n_{1}+n_{2} \\
k_{1}-k_{2} & k_{0}-k_{3} & n_{1}-n_{2} & n_{0}-n_{3} \\
l_{0}+l_{3} & l_{1}+l_{2} & m_{0}+m_{3} & m_{1}+m_{2} \\
l_{1}-l_{2} & l_{0}-l_{3} & m_{1}-m_{2} & m_{0}-m_{3}%
\end{array}%
\right\vert ,  \notag \\
&&\left.
\begin{array}{rrrr}
k_{0}=\cosh \;\beta , & \;k_{1}=0, & \;k_{2}=0, & \;k_{3}=0, \\
m_{0}=\cosh \;\beta , & \;m_{1}=0, & \;m_{2}=0, & \;m_{3}=0, \\
n_{0}=0, & \;n_{1}=0, & \;n_{2}=-\sinh \;\beta , & \;n_{3}=0, \\
l_{0}=0, & \;l_{1}=0, & \;l_{2}=-\sinh \;\beta , & \;l_{3}=0.%
\end{array}%
\right.   \label{7.7.2}
\end{eqnarray}%
The restrictions (\ref{2.1a}) give
\begin{equation*}
\cosh \;\beta S_{0}+\sinh \;\beta S_{3}\geq 0\qquad \Longrightarrow \qquad
e^{\beta }(S_{0}+S_{3})+e^{-\beta }(S_{0}-S_{3})\geq 0,
\end{equation*}%
which holds good for any $\beta $. At the same time we have%
\begin{equation*}
\begin{array}{c}
(\cosh \;\beta S_{0}+\sinh \;\beta S_{3})^{2}-(\cosh \;\beta S_{1}-\sinh
\;\beta S_{2})^{2}-\medskip \\
-(-\sinh \;\beta S_{1}+\cosh \;\beta S_{2})^{2}-(\sinh \;\beta S_{0}+\cosh
\;\beta S_{3})^{2}\geq 0\Longrightarrow \medskip \\
\Longrightarrow S_{0}^{2}-S_{3}^{2}-\cosh \;2\beta
(S_{1}^{2}+S_{2}^{2})+2\sinh \;2\beta S_{1}S_{2}\geq 0.%
\end{array}%
\end{equation*}%
In the variables $a,b,c$ and $y=\tanh \;\beta ,$ $y\in (-1,+1)$, the
second inequality takes the form
\begin{equation*}
1-c^{2}-{\frac{1+y^{2}}{1-y^{2}}}(a^{2}+b^{2})+{\frac{2y}{1-y^{2}}}2ab\geq 0,
\end{equation*}%
or
\begin{equation*}
-y^{2}(a^{2}+b^{2}+1-c^{2})+4aby+(1-a^{2}-b^{2}-c^{2})\geq 0.
\end{equation*}%
The coefficient of $y^{2}$ is positive, so the inequality holds good in the
interval between the roots of the corresponding quadratic equation, that is%
\begin{eqnarray}
y &\in &[y_{1},y_{2}],\qquad y_{1}\leq \tanh \;\beta \leq u_{2},  \notag
\\
y_{1} &=&{\frac{2ab-\sqrt{%
4a^{2}b^{2}+(1-a^{2}-b^{2}-c^{2})(a^{2}+b^{2}+1-c^{2})}}{a^{2}+b^{2}+1-c^{2}}%
}<0,  \notag \\
y_{2} &=&{\frac{2ab+\sqrt{%
4a^{2}b^{2}+(1-a^{2}-b^{2}-c^{2})(a^{2}+b^{2}+1-c^{2})}}{a^{2}+b^{2}+1-c^{2}}%
}>0.  \label{7.7.5}
\end{eqnarray}

Let us check the necessary condition $y_{2}\leq +1$; we have%
\begin{equation*}
\sqrt{4a^{2}b^{2}+(1-a^{2}-b^{2}-c^{2})(a^{2}+b^{2}+1-c^{2})}\leq
(a^{2}+b^{2}+1-c^{2})-2ab,
\end{equation*}%
which after being squared, leads to%
\begin{equation*}
\begin{array}{c}
4a^{2}b^{2}+(1-a^{2}-b^{2}-c^{2})(a^{2}+b^{2}+1-c^{2})\leq \medskip \\
\leq (a^{2}+b^{2}+1-c^{2})^{2}+4a^{2}b^{2}-4ab(a^{2}+b^{2}+1-c^{2}).%
\end{array}%
\end{equation*}
This is equivalent to
\begin{equation*}
0\leq 2(a^{2}+b^{2})-4ab\qquad \Longleftrightarrow \qquad 0\leq (a-b)^{2}.
\end{equation*}%
Therefore, the relation $y_{2}\leq +1$ is satisfied.

\smallskip Now let us verify the second necessary condition $y_{1}\geq -1$:
we have%
\begin{equation*}
-\sqrt{4a^{2}b^{2}-(1-a^{2}-b^{2}-c^{2})(a^{2}+b^{2}+1-c^{2})}\geq
-(a^{2}+b^{2}+1-c^{2})-2ab,
\end{equation*}%
which yields%
\begin{equation*}
\begin{array}{c}
4a^{2}b^{2}+(1-a^{2}-b^{2}-c^{2})(a^{2}+b^{2}+1-c^{2})\leq \medskip \\
\leq (a^{2}+b^{2}+1-c^{2})^{2}+4a^{2}b^{2}-4ab(a^{2}+b^{2}+1-c^{2}).%
\end{array}%
\end{equation*}
This is equivalent to
\begin{equation*}
0\leq 2(a^{2}+b^{2})-4ab\;\Longleftrightarrow \;0\leq (a-b)^{2},
\end{equation*}%
and hence the relation $y_{1}\geq -1$ holds good.

For the case of completely polarized initial light, we have $%
a^{2}+b^{2}+c^{2}=1$ and the formulas become much simpler:
\begin{eqnarray}
y &\in &[y_{1},y_{2}],\qquad y_{1}\leq \tanh \;\beta \leq y_{2},  \notag
\\
y_{1} &=&{\frac{ab-\sqrt{a^{2}b^{2}}}{a^{2}+b^{2}}},\qquad y_{2}={\frac{ab+%
\sqrt{a^{2}b^{2}}}{a^{2}+b^{2}}},  \notag
\end{eqnarray}%
so that
\begin{eqnarray}
ab &>&0,\qquad y_{1}=0,\qquad y_{2}={\frac{2ab}{a^{2}+b^{2}}}\leq +1;  \notag
\\
ab &<&0,\qquad y_{1}={\frac{2ab}{a^{2}+b^{2}}}\geq -1,\qquad y_{2}=0.
\label{7.7.6}
\end{eqnarray}

\vspace{5mm}

\textbf{(7.8)}

\vspace{5mm}

\underline{Variant $U_{3}^{A}(i\beta )$}
\begin{equation*}
U_{3}^{A}(i\beta )=\left\vert
\begin{array}{cccc}
\cosh \;\beta  & 0 & -\sinh \;\beta  & 0 \\
0 & \cosh \;\beta  & 0 & -\sinh \;\beta  \\
-\sinh \;\beta  & 0 & \cosh \;\beta  & 0 \\
0 & -\sinh \;\beta  & 0 & \cosh \;\beta
\end{array}%
\right\vert ,
\end{equation*}%
or in block form
\begin{eqnarray}
M &=&\left\vert
\begin{array}{cccc}
k_{0}+k_{3} & k_{1}+k_{2} & n_{0}+n_{3} & n_{1}+n_{2} \\
k_{1}-k_{2} & k_{0}-k_{3} & n_{1}-n_{2} & n_{0}-n_{3} \\
l_{0}+l_{3} & l_{1}+l_{2} & m_{0}+m_{3} & m_{1}+m_{2} \\
l_{1}-l_{2} & l_{0}-l_{3} & m_{1}-m_{2} & m_{0}-m_{3}%
\end{array}%
\right\vert ,  \notag \\
&&\left.
\begin{array}{rrrr}
k_{0}=\cosh \;\beta , & \;k_{1}=0, & \;k_{2}=0, & \;k_{3}=0, \\
m_{0}=\cosh \;\beta , & \;m_{1}=0, & \;m_{2}=0, & \;m_{3}=0, \\
n_{0}=-\sinh \;\beta , & \;n_{1}=0, & \;n_{2}=0, & \;n_{3}=0, \\
l_{0}=-\sinh \;\beta , & \;l_{1}=0, & \;l_{2}=0, & \;l_{3}=0.%
\end{array}%
\right.   \label{7.8.2}
\end{eqnarray}%
The restrictions (\ref{2.1a}) lead to%
\begin{equation*}
\begin{array}{c}
\cosh \;\beta S_{0}-\sinh \;\beta S_{2}\geq 0\qquad \mbox{holds good for all
}\;\beta ,\medskip \\
(\cosh \;\beta S_{0}-\sinh \;\beta S_{2})^{2}-(\cosh \;\beta S_{1}-\sinh
\;\beta S_{3})^{2}-\medskip \\
-(-\sinh \;\beta S_{0}+\cosh \;\beta S_{2})^{2}-(-\sinh \;\beta S_{1}+\cosh
\;\beta S_{3})^{2}\geq 0\Longrightarrow \medskip \\
\Longrightarrow S_{0}^{2}-S_{2}^{2}-\cosh \;2\beta
(S_{1}^{2}+S_{3}^{2})+2\sinh \;2\beta S_{1}S_{3}\geq 0.%
\end{array}%
\end{equation*}
In the variables $a,b,c$ the second inequality becomes
\begin{equation*}
1-b^{2}-\cosh \;2\beta (a^{2}+c^{2})+\sinh \;2\beta \;2ac\geq 0.
\end{equation*}%
This differs from (\ref{7.1.3}) only by notations, so the results will be
much the same.

\vspace{5mm}

\textbf{(7.9)}

\vspace{5mm}

\underline{Variant $U_{1}^{B}(i\beta )$}
\begin{equation*}
U_{1}^{B}(i\beta )=\left\vert
\begin{array}{cccc}
\cosh \;\beta  & 0 & 0 & -\sinh \;\beta  \\
0 & \cosh \;\beta  & -\sinh \;\beta  & 0 \\
0 & -\sinh \;\beta  & \cosh \;\beta  & 0 \\
-\sinh \;\beta  & 0 & 0 & \cosh \;\beta
\end{array}%
\right\vert ,
\end{equation*}%
or in block form
\begin{eqnarray}
M &=&\left\vert
\begin{array}{cccc}
k_{0}+k_{3} & k_{1}+k_{2} & n_{0}+n_{3} & n_{1}+n_{2} \\
k_{1}-k_{2} & k_{0}-k_{3} & n_{1}-n_{2} & n_{0}-n_{3} \\
l_{0}+l_{3} & l_{1}+l_{2} & m_{0}+m_{3} & m_{1}+m_{2} \\
l_{1}-l_{2} & l_{0}-l_{3} & m_{1}-m_{2} & m_{0}-m_{3}%
\end{array}%
\right\vert ,  \notag \\
&&\left.
\begin{array}{rrrr}
k_{0}=\cosh \;\beta , & \;k_{1}=0, & \;k_{2}=0, & \;k_{3}=0, \\
m_{0}=\cosh \;\beta , & \;m_{1}=0, & \;m_{2}=0, & \;m_{3}=0, \\
n_{0}=0, & \;n_{1}=-\sinh \;\beta , & \;n_{2}=0, & \;n_{3}=0, \\
l_{0}=0, & \;l_{1}=-\sinh \;\beta , & \;l_{2}=0, & \;l_{3}=0.%
\end{array}%
\right.   \label{7.9.2}
\end{eqnarray}%
The restrictions (\ref{2.1a}) give
\begin{equation*}
\cosh \;\beta S_{0}-\sinh \;\beta S_{3}\geq 0,
\end{equation*}%
which holds good for all $\beta $. At the same time we have%
\begin{equation*}
\begin{array}{c}
(\cosh \;\beta S_{0}-\sinh \;\beta S_{3})^{2}-(\cosh \;\beta S_{1}-\sinh
\;\beta S_{2})^{2}-\medskip \\
-(-\sinh \;\beta S_{1}+\cosh \;\beta S_{2})^{2}-(-\sinh \;\beta S_{0}+\cosh
\;\beta S_{3})^{2}\geq 0\Longrightarrow \medskip \\
\Longrightarrow S_{0}^{2}-S_{3}^{2}-\cosh \;2\beta
(S_{1}^{2}+S_{2}^{2})+2\sinh \;2\beta S_{1}S_{2}\geq 0.%
\end{array}%
\end{equation*}

In the variables $a,b,c$ the second inequality becomes
\begin{equation*}
1-c^{2}-\cosh \;2\beta (a^{2}+b^{2})+\sinh \;2\beta \;2ab\geq 0.
\end{equation*}%
This differs from (\ref{7.1.3}) only in notation, so the results will be
much the same.

\vspace{5mm}

\textbf{(7.10)}

\vspace{5mm}

\underline{Variant $U_{3}^{B}(i\beta )$ }
\begin{equation*}
U_{3}^{B}(i\beta )=\left\vert
\begin{array}{cccc}
\cosh \;\beta  & \sinh \;\beta  & 0 & 0 \\
\sinh \;\beta  & \cosh \;\beta  & 0 & 0 \\
0 & 0 & \cosh \;\beta  & -\sinh \;\beta  \\
0 & 0 & -\sinh \;\beta  & \cosh \;\beta
\end{array}%
\right\vert ,
\end{equation*}%
or in block form
\begin{eqnarray}
M &=&\left\vert
\begin{array}{cccc}
k_{0}+k_{3} & k_{1}+k_{2} & n_{0}+n_{3} & n_{1}+n_{2} \\
k_{1}-k_{2} & k_{0}-k_{3} & n_{1}-n_{2} & n_{0}-n_{3} \\
l_{0}+l_{3} & l_{1}+l_{2} & m_{0}+m_{3} & m_{1}+m_{2} \\
l_{1}-l_{2} & l_{0}-l_{3} & m_{1}-m_{2} & m_{0}-m_{3}%
\end{array}%
\right\vert ,  \notag \\
&&\left.
\begin{array}{rrrr}
k_{0}=\cosh \;\beta , & \;k_{1}=\sinh \;\beta , & \;k_{2}=0, & \;k_{3}=0, \\
m_{0}=\cosh \;\beta , & \;m_{1}=-\sinh \;\beta , & \;m_{2}=0, & \;m_{3}=0,
\\
n_{0}=0, & \;n_{1}=0, & \;n_{2}=0, & \;n_{3}=0, \\
l_{0}=0, & \;l_{1}=0, & \;l_{2}=0, & \;l_{3}=0.%
\end{array}%
\right.   \label{7.10.2}
\end{eqnarray}%
The restrictions (\ref{2.1a}) give
\begin{equation*}
\cosh \;\beta S_{0}+\sinh \;\beta S_{1}\geq 0,
\end{equation*}%
which holds good for all $\beta $. At the same time we have%
\begin{equation*}
\begin{array}{c}
(\cosh \;\beta S_{0}+\sinh \;\beta S_{1})^{2}-(\sinh \;\beta S_{0}+\cosh
\;\beta S_{1})^{2}-\medskip \\
-(\cosh \;\beta S_{2}-\sinh \;\beta S_{3})^{2}-(-\sinh \;\beta S_{2}+\cosh
\;\beta S_{3})^{2}\geq 0\Longrightarrow \medskip \\
\Longrightarrow S_{0}^{2}-S_{1}^{2}-\cosh \;2\beta
(S_{2}^{2}+S_{3}^{2})+2\sinh \;2\beta S_{2}S_{3}\geq 0.%
\end{array}%
\end{equation*}%
In the variables $a,b,c$ the second inequality becomes
\begin{equation*}
1-a^{2}-\cosh \;2\beta (b^{2}+c^{2})+\sinh \;2\beta \;2bc\geq 0.
\end{equation*}%
This differs from (\ref{7.1.3}) only in notation, so the results will be
much the same.

\vspace{5mm}

\textbf{(7.11)}

\vspace{5mm}

\underline{Variant $U_{1}^{C}(i\beta )$}
\begin{equation*}
U_{1}^{C}(i\beta )=\left\vert
\begin{array}{cccc}
\cosh \;\beta  & 0 & \sinh \;\beta  & 0 \\
0 & \cosh \;\beta  & 0 & -\sinh \;\beta  \\
\sinh \;\beta  & 0 & \cosh \;\beta  & 0 \\
0 & -\sinh \;\beta  & 0 & \cosh \;\beta
\end{array}%
\right\vert ,
\end{equation*}%
or in block form
\begin{eqnarray}
M &=&\left\vert
\begin{array}{cccc}
k_{0}+k_{3} & k_{1}+k_{2} & n_{0}+n_{3} & n_{1}+n_{2} \\
k_{1}-k_{2} & k_{0}-k_{3} & n_{1}-n_{2} & n_{0}-n_{3} \\
l_{0}+l_{3} & l_{1}+l_{2} & m_{0}+m_{3} & m_{1}+m_{2} \\
l_{1}-l_{2} & l_{0}-l_{3} & m_{1}-m_{2} & m_{0}-m_{3}%
\end{array}%
\right\vert ,  \notag \\
&&\left.
\begin{array}{rrrr}
k_{0}=\cosh \;\beta , & \;k_{1}=0, & \;k_{2}=0, & \;k_{3}=0, \\
m_{0}=\cosh \;\beta , & \;m_{1}=0, & \;m_{2}=0, & \;m_{3}=0, \\
n_{0}=0, & \;n_{1}=0, & \;n_{2}=0, & \;n_{3}=\sinh \;\beta , \\
l_{0}=0, & \;l_{1}=0, & \;l_{2}=0, & \;l_{3}=\sinh \;\beta .%
\end{array}%
\right.   \label{7.11.2}
\end{eqnarray}%
The restrictions (\ref{2.1a}) give
\begin{equation*}
\cosh \;\beta S_{0}+\sinh \;\beta S_{2}\geq 0,
\end{equation*}%
which holds good for all $\beta $. We also have%
\begin{equation*}
\begin{array}{c}
(\cosh \;\beta S_{0}+\sinh \;\beta S_{2})^{2}-(\cosh \;\beta S_{1}-\sinh
\;\beta S_{3})^{2}-\medskip \\
-(\sinh \;\beta S_{0}+\cosh \;\beta S_{2})^{2}-(-\sinh \;\beta S_{1}+\cosh
\;\beta S_{3})^{2}\geq 0\Longrightarrow \medskip \\
\Longrightarrow S_{0}^{2}-S_{2}^{2}-\cosh \;2\beta
(S_{1}^{2}+S_{3}^{2})+2\sinh \;2\beta S_{1}S_{3}\geq 0.%
\end{array}%
\end{equation*}%
In the variables $a,b,c$ the second inequality becomes
\begin{eqnarray}
e^{\beta }(1+b)+e^{-\beta }(1-b) &\geq &0\Leftrightarrow   \notag \\
1-b^{2}-\cosh \;2\beta (a^{2}+c^{2})+\sinh \;2\beta \;2ac &\geq &0.
\label{7.11.3}
\end{eqnarray}%
This differs from (\ref{7.1.3}) only by notation, so the results will be
much the same.

\vspace{5mm}

\textbf{(7.12)}

\vspace{5mm}

\underline{Variant $U_{2}^{C}(-i\beta )$}
\begin{equation*}
U_{2}^{C}(-i\beta )=\left\vert
\begin{array}{cccc}
\cosh \;\beta  & -\sinh \;\beta  & 0 & 0 \\
-\sinh \;\beta  & \cosh \;\beta  & 0 & 0 \\
0 & 0 & \cosh \;\beta  & -\sinh \;\beta  \\
0 & 0 & -\sinh \;\beta  & \cosh \;\beta
\end{array}%
\right\vert ,
\end{equation*}%
or in block form
\begin{eqnarray}
M &=&\left\vert
\begin{array}{cccc}
k_{0}+k_{3} & k_{1}+k_{2} & n_{0}+n_{3} & n_{1}+n_{2} \\
k_{1}-k_{2} & k_{0}-k_{3} & n_{1}-n_{2} & n_{0}-n_{3} \\
l_{0}+l_{3} & l_{1}+l_{2} & m_{0}+m_{3} & m_{1}+m_{2} \\
l_{1}-l_{2} & l_{0}-l_{3} & m_{1}-m_{2} & m_{0}-m_{3}%
\end{array}%
\right\vert ,  \notag \\
&&\left.
\begin{array}{rrrr}
k_{0}=\cosh \;\beta , & \;k_{1}=-\sinh \;\beta , & \;k_{2}=0, & \;k_{3}=0,
\\
m_{0}=\cosh \;\beta , & \;m_{1}=-\sinh \;\beta , & \;m_{2}=0, & \;m_{3}=0,
\\
n_{0}=0, & \;n_{1}=0, & \;n_{2}=0, & \;n_{3}=0, \\
l_{0}=0, & \;l_{1}=0, & \;l_{2}=0, & \;l_{3}=0.%
\end{array}%
\right.   \label{7.12.2}
\end{eqnarray}%
The restrictions (\ref{2.1a}) give
\begin{equation*}
\cosh \;\beta S_{0}-\sinh \;\beta S_{1}\geq 0\;,
\end{equation*}%
which holds good for all $\beta $. We also have%
\begin{equation*}
\begin{array}{c}
(\cosh \;\beta S_{0}-\sinh \;\beta S_{1})^{2}-(-\sinh \;\beta S_{0}+\cosh
\;\beta S_{1})^{2}-\medskip \\
-(\cosh \;\beta S_{2}-\sinh \;\beta S_{3})^{2}-(-\sinh \;\beta S_{2}+\cosh
\;\beta S_{3})^{2}\geq 0\Longrightarrow \medskip \\
\Longrightarrow S_{0}^{2}-S_{1}^{2}-\cosh \;2\beta
(S_{2}^{2}+S_{3}^{2})+2\sinh \;2\beta S_{2}S_{3}\geq 0.%
\end{array}%
\end{equation*}%
In the variables $a,b,c$ the second inequality becomes
\begin{equation*}
1-a^{2}-\cosh \;2\beta (b^{2}+c^{2})+\sinh \;2\beta \;2bc\geq 0.
\end{equation*}%
It differs from (\ref{7.1.3}) only in notation, so the results will be much
the same.

%=========================================================

\section{Varying the degree of polarization of the light}

Let us examine how the degree of polarization of the light changes for the
above 12 special one-parametric Mueller transformations. The corresponding
characteristics may be given by the difference
\begin{equation*}
(a^{^{\prime }2}+b^{^{\prime }2}+c^{^{\prime }2})-(a^{2}+b^{2}+c^{2})=D.
\end{equation*}%
The $D$-entities for the first six cases \textbf{(7.1)--(7.6)} are
\begin{eqnarray}
(7.1)\qquad D &=&{\frac{(a-x)^{2}+(b^{2}+c^{2})(1+x^{2})}{(1+ax)^{2}}}%
-a^{2}-b^{2}-c^{2},  \notag \\
(7.2)\qquad D &=&{\frac{(b-x)^{2}+(a^{2}+c^{2})(1+x^{2})}{(1+bx)^{2}}}%
-a^{2}-b^{2}-c^{2},  \notag \\
(7.3)\qquad D &=&{\frac{(c-x)^{2}+(a^{2}+b^{2})(1+x^{2})}{(1+ac)^{2}}}%
-a^{2}-b^{2}-c^{2},  \notag \\
(7.4)\qquad D &=&{\frac{(a-x)^{2}+(b^{2}+c^{2})(1+x^{2})}{(1+ax)^{2}}}%
-a^{2}-b^{2}-c^{2},  \notag \\
(7.5)\qquad D &=&{\frac{(b-x)^{2}+(a^{2}+c^{2})(1+x^{2})}{(1+bx)^{2}}}%
-a^{2}-b^{2}-c^{2},  \notag \\
(7.6)\qquad D &=&{\frac{(c-x)^{2}+(a^{2}+b^{2})(1+x^{2})}{(1+ac)^{2}}}%
-a^{2}-b^{2}-c^{2}.  \label{8.2}
\end{eqnarray}%
It will be sufficient to consider in detail only one case, e.g., \textbf{%
(7.1)}. In this case, we get:
\begin{equation*}
D={\frac{(a-x)^{2}}{(1+xa)^{2}}}+{\frac{(b^{2}+c^{2})(x^{2}+1)}{(1+xa)^{2}}}%
-(a^{2}+b^{2}+c^{2}).
\end{equation*}%
The degree of polarization decreases\footnote{%
We use the constraint \ $a^{2}+b^{2}+c^{2}=p^{2}.$} when
\begin{equation*}
D<0,\qquad (a-x)^{2}+(b^{2}+c^{2})(x^{2}+1)<p^{2}(1+xa)^{2},
\end{equation*}%
or when
\begin{equation*}
x^{2}(1+b^{2}+c^{2})-2ax<x^{2}a^{2}p^{2}+2xap^{2},
\end{equation*}%
and this yields
\begin{equation*}
D<0,\qquad x^{2}(1-a^{2})-2ax<0.
\end{equation*}%
Thus, the light is depolarized if
\begin{eqnarray}
D &<&0,\qquad 0<x<{\frac{2a}{1-a^{2}}}\;,\qquad a>0;  \notag \\
D &<&0,\qquad {\frac{2a}{1-a^{2}}}<x<0\;,\qquad a<0.  \label{8.4b}
\end{eqnarray}%
In contrast, the degree of polarization of the light increases if
\begin{equation}
x^{2}(1-a^{2})-2ax>0,  \label{8.5a}
\end{equation}%
that is, when
\begin{eqnarray}
D &>&0,\qquad x>{\frac{2a}{1-a^{2}}}\;,\qquad a>0;  \notag \\
D &>&0,\qquad x<{\frac{2a}{1-a^{2}}}\;,\qquad a<0.  \label{8.5b}
\end{eqnarray}%
The light does not change the degree of polarization if
\begin{equation*}
D=0\qquad \Longrightarrow \qquad x={\frac{2a}{1-a^{2}}}.
\end{equation*}%
This fact is illustrated by Fig. 2.

\begin{figure}[h!]
\centering
\includegraphics[width=12cm]{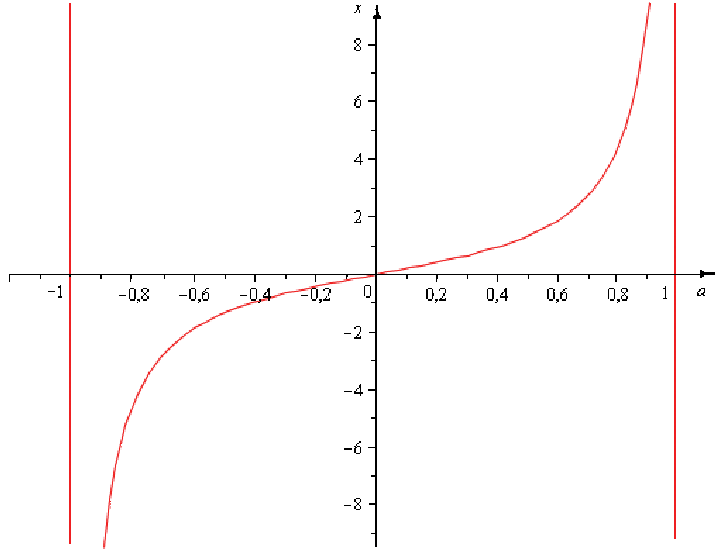}
\par
\vspace{-20mm}
\caption{The line $D(a,x)=0$}
\end{figure}

Now let us examine the other six cases \textbf{(7.7)--(7.12)}.

The case \textbf{(7.7)} (we use the notation $\tanh \;\beta =y$) gives
\begin{equation}
\begin{array}{c}
D={\dfrac{(a\;\cosh \;\beta -b\;\sinh \;\beta )^{2}+(-a\;\sinh \;\beta
+b\;\cosh \;\beta )^{2}+(\sinh \;\beta +c\;\cosh \;\beta )^{2}}{(\cosh
\;\beta +c\;\sinh \;\beta )^{2}}}-\medskip \\
-a^{2}-b^{2}-c^{2}={\dfrac{(a-by)^{2}+(b-ay)^{2}+(c+y)^{2}}{(1+cy)^{2}}}%
-a^{2}-b^{2}-c^{2}.%
\end{array}
\label{8.7}
\end{equation}%
The degree of polarization decreases when
\begin{equation*}
D<0,\qquad (a-by)^{2}+(b-ay)^{2}+(c+y)^{2}<(a^{2}+b^{2}+c^{2})(1+cy)^{2},
\end{equation*}%
or when
\begin{equation*}
y^{2}(a^{2}+b^{2}+1)+2y(c-2ab)<y^{2}c^{2}p^{2}+2ycp^{2},
\end{equation*}%
and this leads us to
\begin{equation*}
D<0\;,\qquad y^{2}(1-c^{2})(1+p^{2})-2y[2ab-c(1-p^{2})]<0.
\end{equation*}%
The solutions have the form
\begin{eqnarray}
D &<&0,\qquad 2ab-c(1-p^{2})>0,\qquad 0<y<{\frac{2}{(1-c^{2})(1+p^{2})}}\;%
\left[ 2ab-c(1-p^{2})\right] ,  \notag \\
D &<&0,\qquad 2ab-c(1-p^{2})<0,\qquad {\frac{2}{(1-c^{2})(1+p^{2})}}\;\left[
2ab-c(1-p^{2})\right] <y<0.  \label{8.8b}
\end{eqnarray}

The degree of polarization increases when
\begin{equation*}
D>0\;,\qquad (a-by)^{2}+(b-ay)^{2}+(c+y)^{2}>(a^{2}+b^{2}+c^{2})(1+cy)^{2}.
\end{equation*}%
This gives
\begin{equation*}
y^{2}(1-c^{2})(1+p^{2})-2y[2ab-c(1-p^{2})]>0.
\end{equation*}%
The solutions are
\begin{eqnarray}
D &>&0,\qquad 2ab-c(1-p^{2})>0,\qquad y>{\frac{2}{(1-c^{2})(1+p^{2})}}\;%
\left[ 2ab-c(1-p^{2})\right] ,  \notag \\
D &>&0,\qquad 2ab-c(1-p^{2})<0,\qquad y<{\frac{2}{(1-c^{2})(1+p^{2})}}\;%
\left[ 2ab-c(1-p^{2})\right] .  \label{8.9b}
\end{eqnarray}%
The light does not change the degree of polarization if
\begin{equation*}
D=0,\qquad y={\frac{2}{(1-c^{2})(1+p^{2})}}\;\left[ 2ab-c(1-p^{2})\right] .
\end{equation*}

The formulas become considerably simpler when the initial light is completely
polarized. \linebreak
All the six variants of this type are characterized by the
following $D$-entities:
\begin{eqnarray}
(7.7)\qquad D &=&{\frac{(a-by)^{2}+(b-ay)^{2}+(y+c)^{2}}{(1+cy)^{2}}}%
-a^{2}-b^{2}-c^{2},  \notag \\
(7.9)\qquad D &=&{\frac{(a-by)^{2}+(b-ay)^{2}+(y-c)^{2}}{(1-cy)^{2}}}%
-a^{2}-b^{2}-c^{2},  \notag \\
(7.8)\qquad D &=&{\frac{(a-cy)^{2}+(c-ay)^{2}+(y-b)^{2}}{(1-by)^{2}}}%
-a^{2}-b^{2}-c^{2},  \notag \\
(7.11)\qquad D &=&{\frac{(a-cy)^{2}+(c-ay)^{2}+(y+b)^{2}}{(1+by)^{2}}}%
-a^{2}-b^{2}-c^{2},  \notag \\
(7.10)\qquad D &=&{\frac{(b-cy)^{2}+(c-by)^{2}+(y+a)^{2}}{(1+ay)^{2}}}%
-a^{2}-b^{2}-c^{2},  \notag \\
(7.12)\qquad D &=&{\frac{(b-cy)^{2}+(c-by)^{2}+(y-a)^{2}}{(1-ay)^{2}}}%
-a^{2}-b^{2}-c^{2}.  \label{8.11}
\end{eqnarray}

It is obvious that the differences between these relations consist of
notations only, so the derived results will be similar.

\vspace{5mm}

Examination of other subgroups (see Section 7)  is
a subject for further concern.

\section{ Acknowledgements}

The present work was developed under the auspices of Grant
    1196/2012 - BRFFR - RA No. F12RA-002, within the cooperation framework between Romanian
    Academy and Belarusian Republican Foundation for Fundamental Research.

%=========================================================

\vspace{10mm}

{\bf Elena Ovsiyuk and Olga  Veko}\\
Mosyr State Pedagogical University, Republic of Belarus.\\
E-mail: e.ovsiyuk@mail.ru , vekoolga@mail.ru,\\[1.6mm]

{\bf Mircea Neagu}\\
University Transilvania of Brasov, Romania.\\
E-mail: mirceaneagu73@gmail.com\\[1.6mm]

{\bf Vladimir Balan}\\
University Politehnica of Bucharest, Romania.\\
E-mail: vladimir.balan@upb.ro\\[1.6mm]

{\bf Victor Red'kov}\\
B.I. Stepanov Institute of Physics,\\
National Academy of Sciences of Belarus, Minsk, Republic of Belarus.\\
E-mail: v.redkov@dragon.bas-net.by
%

%=========================================================

\end{document}